\documentclass[twocolumn,showpacs,preprintnumbers,amsmath,amssymb]{revtex4}

\usepackage{dcolumn}
\usepackage{bm}
\usepackage{graphicx}

\def\ba{\begin{eqnarray}}
\def\ea{\end{eqnarray}}
\def\F{\mathcal{F}}
\def\n{\mathbf{n}}
\def\th{\textrm{\mbox{\tiny{th}}}}
\def\Tobs{T_{\textrm{\mbox{\tiny{used}}}}}
\def\Tcoh{\Delta T}
\def\Tsft{T_{\textrm{\mbox{\tiny{sft}}}}}
\def\coh{{\textrm{\mbox{\tiny{coh}}}}}

\def\SSB{\textrm{\mbox{\tiny{ssb}}}}

\def\min{\textrm{\mbox{\tiny{min}}}}
\def\max{\textrm{\mbox{\tiny{max}}}}
\def\sft{\textrm{\mbox{\tiny{sft}}}}
\def\res{\textrm{\mbox{\tiny{res}}}}
\def\RMS{\textrm{\mbox{\tiny{RMS}}}}

\def\ib{{(i)}}
\def\bmath#1{\mbox{\boldmath$#1$\unboldmath}}
\def\bmaths#1{\mbox{\scriptsize\boldmath$#1$\unboldmath}}

\def\lesssim{\mathrel{\hbox{\rlap{\hbox{\lower4pt\hbox{$\sim$}}}\hbox{$<$}}}}
\def\gtrsim{\mathrel{\hbox{\rlap{\hbox{\lower4pt\hbox{$\sim$}}}\hbox{$>$}}}}
\def\alt{\mathrel{\hbox{\rlap{\hbox{\lower4pt\hbox{$\sim$}}}\hbox{$<$}}}}
\def\agt{\mathrel{\hbox{\rlap{\hbox{\lower4pt\hbox{$\sim$}}}\hbox{$>$}}}}

\newcommand {\be}{\begin{equation}}
\newcommand {\ee}{\end{equation}}

\begin{document}

\title{Improved Stack-Slide Searches for Gravitational-Wave Pulsars}

\newcommand*{\AEI}{Max-Planck-Institut f\"ur
    Gravitationsphysik, Albert-Einstein-Institut, Am M\"uhlenberg 1,
    D-14476 Golm, Germany}\affiliation{\AEI}

\author{Curt Cutler}\email{curt.cutler@aei.mpg.de}\affiliation{\AEI}

\author{Iraj Gholami}\email{iraj.gholami@aei.mpg.de}\affiliation{\AEI}

\author{Badri Krishnan}\email{badri.krishnan@aei.mpg.de}\affiliation{\AEI}

\date{\today}

\begin{abstract}

We formulate and optimize a computational search strategy for detecting
gravitational waves from isolated, previously-unknown neutron stars
(that is, neutron stars with unknown sky positions, spin frequencies,
and spin-down parameters).
It is well known that fully coherent searches over the relevant
parameter-space volumes are not computationally feasible, and so more
computationally efficient methods are called for.  The
first step in this direction was taken by Brady\&Creighton (2000), who
proposed and optimized a two-stage, stack-slide search algorithm.  We
generalize and otherwise improve upon the Brady-Creighton scheme in
several ways.  Like Brady\&Creighton, we consider a stack-slide scheme,
but here with an
arbitrary number of semi-coherent stages and with a coherent
follow-up stage at the end.
We find that searches with three semi-coherent stages are significantly
more efficient than two-stage searches (requiring about $2$--$5$ times less
computational power for the same sensitivity) and are only slightly
less efficient than searches with four or more stages.
We calculate the signal-to-noise ratio required for detection, as
a function of computing power and neutron star spin-down-age, using
our optimized searches.
\end{abstract}

\preprint{AEI-2005-104}
\maketitle

\section{Introduction}
\label{sec:intro}

In analyzing data from Earth-based and space-based gravitational-wave (GW)
detectors, we will be computationally limited in performing certain
types of searches--especially searches for long-lived signals described
by several unknown parameters.
For such signals, the number of templates signals required to
discretely cover the parameter space (at useful resolution)
typically increases rapidly as a function of the observation time.
For ground-based detectors, such as LIGO, a well-known example is the
search for nearly periodic GWs from unknown, isolated, rapidly
rotating neutron stars (NSs). We will refer to NSs that are continuously
emitting GWs as ``GW pulsars''.
By ``unknown'', we mean that the GW pulsar's
position on the sky, frequency, and frequency derivatives are all
unknown, and so must be searched over. (The NS could be unknown either
because it is electromagnetically inactive, or because its
electromagnetic emission
does not reach us--e.g., because we do not intersect its radio pulsar beam.)
Brady et al.~\cite{bccs}
showed that straightforward matched-filter searches for unknown
GW pulsars would
be severely computationally limited; for example, searches for young, fast
NSs (NSs with GW frequencies as high as $1\,$kHz and spin-down ages
as short as $40$ yr) would be limited to observation times of order one day.
To address this problem, Brady \& Creighton~\cite{bc}
(henceforth referred to as BC) were the first to
consider hierarchical, multistage, semi-coherent searches for GW pulsars.
Briefly, a semi-coherent search is one where a sequence of
short data stretches are all coherently searched, using
some technique akin to matched filtering,
and then the resulting powers from the different stretches are summed.
The method is only
``semi-coherent'' because powers are added instead of complex amplitudes; i.e.,
information regarding the overall phase of the signal in different
stretches is  discarded. This allows one to use a much coarser
grid on parameter space than would be required in a fully coherent search of
the same data.  BC developed a ``stack-slide''
method for summing the powers along different tracks in the time-frequency
plane, in close analogy to the ``power stacking'' method
(sometimes called the Radon transform) used in radio pulsar searches.
The basic idea of their two-stage search is to
identify a list of ``candidates'' (basically, promising-looking
regions in parameter space) in the first stage, using some
fraction of the
available data, and then to ``follow up'' those candidates using more
data in the second stage. In their scheme, both the first and second
stages are semi-coherent.

In this paper we revisit the problem of constructing efficient, hierarchical
searches for GW pulsars. We build on the BC treatment, but
we also significantly generalize and otherwise improve upon their
work. The most important improvements are that we consider searches
with $n$ semi-coherent stages (not just $2$), with surviving
candidates being winnowed at each stage, and we add on a fully coherent
final stage to verify or debunk any remaining candidates.
We also explicitly account for the unknown polarization of the source,
while this complication was omitted for in BC, for simplicity.
Other important differences between our work and theirs will be highlighted
below.

This paper is organized as follows.  Section \ref{sec:basics} sets up
notation, describes the expected signal from an isolated GW pulsar,
and reviews the stack-slide algorithm.  Our general multistage strategy for
searching through large parameter spaces for GW pulsars,
using a combination of semi-coherent methods and coherent methods,
is explained in Section~\ref{sec:hierarchical}.
Our general search scheme contains a fair number of
free parameters (the number
and duration of the coherently analyzed stretches in each semi-coherent
stage, as well as the coarseness of the discrete grid used to
cover the parameter space of sought-for signals), which we
can adjust to make the search as efficient
as possible. Our general scheme for performing this optimization is
described in  Section~\ref{subsec:optimize}.
Section~\ref{sec:formulae} develops all the formulae
we need to evaluate the computational cost of any of our strategies, for
any desired sensitivity.
More specifically, section~\ref{subsec:metric} reviews the
template-counting formulae developed in BC;
section~\ref{subsec:fd} develops the equations
relating the thresholds that candidates must pass at different stages
(to advance to the following stage) to the false dismissal (FD) rates
at those stages, and hence to the overall sensitivity of the
search;
and section~\ref{subsec:cost}
derives estimates for the dominant computational cost of each
part of the search.
Section~\ref{sec:results} describes our results: the optimal
strategy (within our general scheme) and its sensitivity.
Section~\ref{sec:conclusions} concludes with a summary of our main results
and a discussion of open issues and future work.

\section{Notation and basics}
\label{sec:basics}

\subsection{The signal from a GW pulsar}
\label{subsec:pulsar}

Here we briefly review the expected GW
signal from a spinning neutron star.  Let $x(t)$ be the output of some
detector.  In the absence of any signal, $x(t)$ is
just noise $n(t)$, which we shall assume to be a stationary, Gaussian
stochastic process with zero mean.  In the presence of a signal, we
have
\be
x(t) = n(t) + h(t)
\ee
where the signal $h(t)$ is a deterministic function of time.
We assume that the GW pulsar is isolated and at rest with
respect to us, so that effects due to its motion can be neglected.
(More precisely, we assume these effects can absorbed into an overall
Doppler shift, and so are unobservable.)
Let $t_\SSB$ be
time measured in the Solar System Barycenter (SSB) frame. The form of
$h(t)$ in this frame is a constant-amplitude sinusoid with phase given by
\be \label{eq:phase}
\Phi(t_\SSB) = \Phi_0 + 2\pi f_0 \Delta t_\SSB +
2\pi \sum_{k=1}^{s}\frac{f_k}{(k+1)!}\left( \Delta t_\SSB
\right)^{k+1}
\ee
where $\Delta t_\SSB \equiv t_\SSB - t_\SSB^{(0)}$, with $t_\SSB^{(0)}$
being a fiducial start time; $\Phi_0$, $f_0$ and $f_k$ are respectively the
phase, frequency, and spin-down parameters at the start time, and $s$ is the
number of spin-down parameters that we search over.  Assuming that the
pulsar is isolated and emitting GWs due to a small
deviation from axisymmetry, the waveforms for the two polarizations are
\ba
h_+ &=& \frac{1}{2}h_0(1+\cos^2\iota)\cos\Phi(t)\, ,
\label{eq:waveformplus} \\  h_\times  &=&
h_0\cos\iota \sin\Phi(t) \label{eq:waveformcross}
\ea
where $h_0$ represents the angle-independent amplitude of the wave,
$\iota$ is the angle
between the spin-axis of the pulsar and the direction of the waves'
propagation, and
the frequency $f = \dot\Phi/2\pi $ of the emitted GWs is equal
to twice the rotational frequency of the star.

Let $\mathbf{n}$ be the unit vector pointing from the Solar System
toward the pulsar,  $\mathbf{r}(t)$ be the position of the detector in
the SSB frame, and $\mathbf{v}(t)$ be its velocity with $t$ being the
time in the detector frame.  Ignoring relativistic
corrections~\footnote{Of course an actual search should must take into
  account the so-called 
Einstein and Shapiro delays, but these are unimportant for the question
of how the search is most efficiently {\it organized}, which is the
focus of this paper.}, a wave
reaching the Sun at time $t_\SSB$ will reach the detector at time
\be \label{eq:tssb} t = t_\SSB -
\frac{\mathbf{r}(t)\cdot\mathbf{n}}{c} \,.\ee
As seen from Eqs.~(\ref{eq:phase}) and (\ref{eq:tssb}), to a good
approximation, the instantaneous frequency of the
signal as seen by the detector is given by the familiar Doppler shift
expression
\be \label{eq:doppler}
f(t) = \hat{f}(t) \left( 1 +
\frac{\mathbf{v}(t)\cdot\mathbf{n}}{c}\right)
\ee
where $\hat{f}(t)$ is the instantaneous frequency of the signal in the
SSB frame, and is given by
\be \label{eq:fhat}
\hat{f}(t) = f_0 + \sum_{k=1}^{s}\frac{f_k}{k!}\left( \Delta t_\SSB \right)^{k}\,.
\ee
Eqs.~(\ref{eq:doppler}) and (\ref{eq:fhat}) describe the frequency
modulation of the received signal.
The received signal is also amplitude modulated by the
time-changing antenna pattern of the detector as it is carried along
by the Earth's rotation.
The received signal $h(t)$ is a linear
combination of $h_+$ and $h_\times$:
\be\label{eq:detoutput}
h(t) = F_+(\n,\psi)h_+(t) + F_\times(\n,\psi)h_\times(t)
\ee
where $\psi$ is the polarization angle of the signal, and $F_{+,\times}$ are
the antenna pattern functions.  Due to the motion of the Earth, the $F_{+,\times}$
depend implicitly on time:
\ba F_+(t) &=& a(t)\cos 2\psi + b(t)\sin 2\psi  \\
F_\times (t) &=& b(t)\cos 2\psi - a(t)\sin 2\psi \ea
where the functions $a(t)$ and $b(t)$ are independent of $\psi$.
(In these equations, the angle between the arms of the detector is
taken to be $\pi/2$.)
We refer the reader to \cite{jks} for
explicit expressions for $a(t)$ and $b(t)$.

The modulated frequency is described by the $s+3$ parameters
consisting of
$f_0$ and $\vec{\lambda}:= (\mathbf{n}, \{f_k\}_{k=1\ldots s})$; we
shall often denote the pair $(f_0,\vec{\lambda})$ by the boldface symbol
$\bmath{\lambda}$.  Apart from the parameters $\bmath{\lambda}$, the
waveform (\ref{eq:detoutput})
depends on other parameters: the pulsar's orientation $\iota$,
polarization angle $\psi$, the initial phase $\Phi_0$, and the
amplitude $h_0$.  The optimal matched filter statistic
\cite{jks} for detecting the waveform must, in principle, search over
the entire parameter space $(\bmath{\lambda},\iota,
\psi,\Phi_0,h_0)$.  However, it turns out that the computationally
challenging part of the search involves just the $\mathbf{\lambda}$; the
optimization of over $(\iota,
\psi, \Phi_0, h_0)$ can be done analytically,
by means of the
$\F$-statistic defined in \cite{jks}.  The $\F$-statistic {\it is}
the optimal matched filter statistic maximized over $(\iota,
\psi, \Phi_0, h_0)$.  It is therefore only a function of
$(f_0,\vec{\lambda})$ and it is given by
\be\label{eq:fstatdef}
\F(f_0,\vec{\lambda}) = 4\left[\frac{B|F_a|^2 +
A|F_b|^2 - 2C\mathcal{R}(F_aF_b^\star)}{\Tcoh S_n(f_0) D} \right]
\ee
where $S_n(f)$ is the single-sided
power spectral density of the detector noise $n(t)$, and
\begin{eqnarray}
F_a &=& \int_{-\Tcoh/2}^{\Tcoh/2}
x(t)a(t)e^{-i\Phi(t;\bmaths{\lambda})}\,dt \, ,\label{eq:Fa}\\
F_b &=& \int_{-\Tcoh/2}^{\Tcoh/2}
x(t)b(t)e^{-i\Phi(t;\bmaths{\lambda})}\,dt \, , \label{eq:Fb} \\
A &=& (a||a)\,, \qquad B = (b||b) \,,\\
C &=& (a||b)\,,\qquad D = AB-C^2\,.
\end{eqnarray}
Here we have used the notation
\begin{equation}
(x||y) = \frac{2}{\Tcoh}\int_{-\Tcoh/2}^{\Tcoh/2} x(t)y(t)\,dt\,.
\end{equation}
In cases where the amplitude modulation can be ignored (e.g., for short
data segments, $<<1\,$ day long,  where the
$a(t)$ and $b(t)$ can be approximated as constant), we see that $\F$ is
proportional to the demodulated Fourier transform which matches just
the phase evolution:
\be \label{eq:approxF}
\F \propto |\tilde{X}(f,\vec{\lambda})|^2
\ee
where
\be \label{eq:demodulated}
\tilde{X}(f,\lambda) = \int_{-\Tcoh/2}^{\Tcoh/2}x(t)e^{-i\Phi(t;\bmaths{\lambda})}\,dt\,.
\ee
The $\F$-statistic is the optimal (frequentist) detection
statistic for GW pulsars, and it is at the core of
some algoriths currently used to search for GW pulsars
in  LIGO and GEO data~\cite{S1:pulsar}.
Some important properties of the
$\F$-statistic are reviewed in section IV.B, and
a more detailed description can be found in \cite{jks}.

\subsection{The Stack-Slide algorithm}
\label{subsec:stackslide}

The stack-slide algorithm is best described with reference to the Doppler
shift formula of Eq.~(\ref{eq:doppler}). Imagine we have a data
stream $x(t)$ covering an observation time $\Tobs$, and we wish to
search for a GW pulsar with some parameters $\bmath{\lambda}$.  We break up the
data into $N$ smaller segments of length $\Tcoh=\Tobs/N$, and
calculate the Fourier spectrum of each segment. For now we assume each
segment is sufficiently short that the signal frequency remains
confined to a single discrete frequency bin.  If there is a signal present, it
will most likely be too weak to show up in a single segment with any
significant signal-to-noise-ratio (SNR).  However, we can increase the
SNR by adding the power from the different segments.  We must {\it not}
use the the same frequency bin from each segment,
but rather must follow
the frequency evolution given by Eq.~(\ref{eq:doppler}).  Thus,
we `stack' the power after `sliding' each segment in frequency space.
Note that the sliding depends on $\vec{\lambda}$.  Thus, in practice, we
choose a grid in the space of $\vec{\lambda}$'s and the sliding is done
differently at each grid point.

As described above, the sensitivity of the stack-slide algorithm is
restricted due
to the length of $\Tcoh$; we should not take $\Tcoh$ to be too large, since then
we would lose SNR due to the signal power being spread over several frequency bins.
However, we can gather all the signal power back into a
single bin by taking account of the Doppler modulation and spin-down while
calculating the spectrum of a segment; i.e., we de-modulate each data
segment before summing.

With these concepts at hand, we can now describe the stack-slide
search for the $\F$-statistic.   The strategy is very similar to the
power summing method described earlier in this section.  Again we
break up the data of length $\Tobs$ into $N$ segments, each of length
$\Tcoh = \Tobs/N$.  We choose a point $\vec{\lambda}_d$ in
parameter space, and demodulate the signal accordingly.  We calculate
$\F(f,\vec{\lambda}_d)$ as a function of the frequency for each
segment and add the $\F$-statistic values after sliding the different
segments in frequency space appropriately.

As explained in BC, the resolution of sky- and
spin-down-space that suffices for the demodulation is not
fine enough for for the stack-slide step. Thus at each stage
we two grids on parameter space: a coarse one for
performing the short-segment demodulations and a fine one for
sliding and stacking the short-segment results.
We refer the reader to \cite{ip}
and the appendix of \cite{hough} for a detailed
derivation of the formula relating the required amount of sliding
to the parameters $\vec{\lambda}_d$.

\section{A multistage hierarchical search}
\label{sec:hierarchical}

\subsection{The general algorithm}
\label{subsec:algorithm}

The stack-slide search algorithm described in the previous section has
two components: 1) calculation of the $\F$-statistic for data stretches of length
$\Tcoh$, 2) summation of the resulting $\F$ values along the appropriate tracks in the time-frequency plane.
(If there are $N$ coherently analyzed segments, then the sums have  $N$ terms.)
If we had unlimited computational resources, we would
simply do a fully coherent search on all the data; i.e., set $N=1$ and take $\Tcoh$
to be the entire observation time.
However, the number of templates required for a fully
coherent search increases as a high power of $\Tcoh$, making this
impractical for all-sky searches.

To illustrate this point, consider an all-sky search for young, fast
pulsars, i.e., GW pulsars that have a spin-down age as short as $\tau_\min
= 40\,$yr and that emit GWs with frequency up to
$f_\max = 1000\,$Hz.
Let us assume that we have $30\,$days of data available to
us.  Imagine two different ways of looking for this pulsar: a full
$30-$day coherent integration versus a semi-coherent method where the
available data is broken up into 30 equal segments.  The formula for
the number of templates required for these searches is given below in
Eq~(\ref{eq:Np}).  It turns out that the full coherent search
requires $\sim 4.2\times 10^{15}$ templates if we are to not lose more
than $30\%$ of the signal power.  On the other hand, the semi-coherent
search requires only $\sim 1.5\times 10^{11}$ templates for the same
allowed fractional loss in signal power.  The ratio of the
the number of templates required for the two types of searches
increases rapidly with the observation time; for instance, for an
observation time of $40\,$ days, the corresponding numbers are $\sim 5.5 \times
10^{16}$ and $\sim 8.3 \times 10^{11}$ for the full coherent and
semi-coherent searches respectively.

\begin{figure}
  \begin{center}
  \includegraphics[width=\columnwidth]{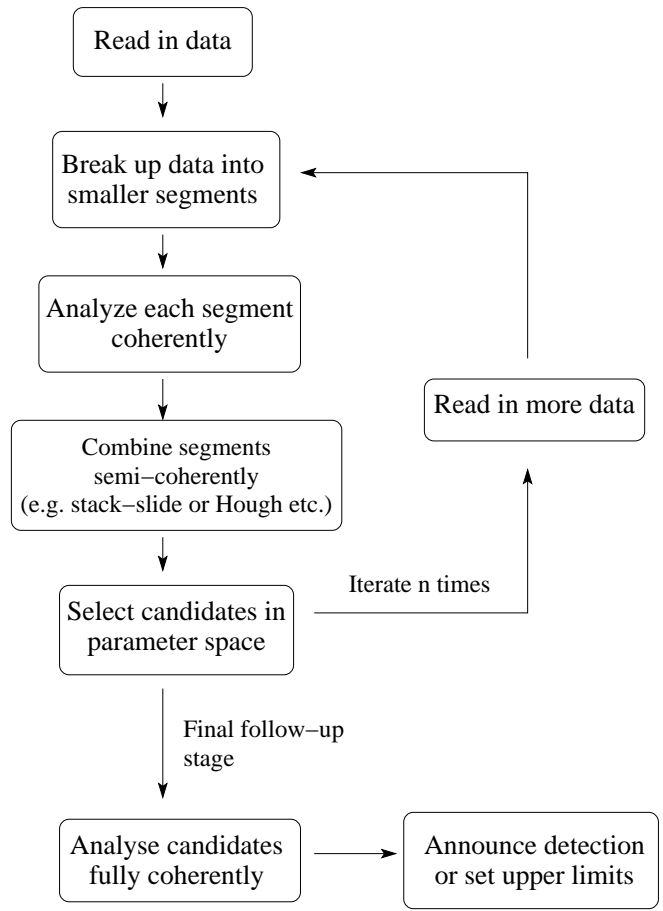}
  \caption{A hierarchical scheme for the analysis of large parameter space
  volumes for continuous wave searches.  Each step analyzes only those
  regions in parameter space that have not been discarded by any of
  the previous steps.}\label{fig:hierarchical}
  \end{center}
\end{figure}

As illustrated by the above example, semi-coherent searches for unknown
GW pulsars are a compromise
forced upon us by limited computing power. Such searches will remain
computationally limited for the foreseeable future, so it behooves us to
organize them as efficiently as possible. In this paper we consider
a class of multistage, hierarchical  search algorithms.
Since our main ``problem'' is the
large volume of parameter space we need to search over,
the basic idea behind these algorithms is to identify and  discard
unpromising regions of parameter space
as fast as possible--without discarding real signals.
The type of scheme we consider is illustrated schematically in
Fig.~\ref{fig:hierarchical}.
The first stage is a semi-coherent search through some fraction of the
available data. A threshold is set, and candidates exceeding
this threshold are passed to the next stage.
The second stage is similar to the
first, but includes additional data and generally entails a finer
resolution of parameter space.
(The latter means that any candidate that survives the first semi-coherent
stage gives rise to a little crowd of nearby candidates that are
examined in the second semi-coherent stage.)
 Any candidate that exceeds the
second-stage threshold is passed on to the third stage, and so on.
In an $(n+1)$-stage search, any candidate surviving all $n$ semi-coherent
selections is subjected to a final, coherent search (which we consider
the $(n+1)^{\rm th}$ stage); if the final,
coherent threshold is exceeded, then a detection is announced.
We impose as a constraint that the false alarm (FA) rate for the entire search
must be $< 1\%$; i.e., if the data is actually just noise, then
the probability that a detection is announced must be $< 1\%$.
For reasons explained below, in realistic examples
this inequality is all too easy to satisfy ;
the actual FA rate for our optimized searches
is typically smaller than $1\%$ by many orders of magnitude.

In the end, our search will be able to detect
a GW pulsar signal whose rms strength (at the detector) $h_\RMS$
exceeds some threshold value $h_{\th}$
(with a false dismissal rate of $10--15\%$).
We can think of $1/h_{\th}$
as the search's sensitivity.
We will optimize our search to get the maximum
sensitivity for any given computing power or, equivalently, to
find the minimum computer power necessary to attain any
given sensitivity.

The problem of optimizing a semi-coherent, hierarchical search
scheme for GW pulsars was first studied by BC.  The
present study builds upon the BC formalism, but there are also
some important differences. We call attention to the following ones:
\begin{description}
\item[1)] BC consider a hierarchical search
consisting of exactly two semi-coherent stages.
In the present work, we consider a
search consisting of an arbitrary number of semi-coherent steps, plus a
fully coherent, ``follow-up'' stage (utilizing all the available data)
to assess the significance of any surviving candidates.
The effect of the final, follow-up stage is to ensure
that the overall false alarm rate (fixed at exactly $1\%$ by BC) is
greatly reduced and, for all practical purposes, ceases to be a
constraint.
\item[2)] In BC's second semi-coherent
stage, all the data used in the first-stage is reanalyzed, along with some
``fresh'' (as yet unanalyzed) data. {\it A priori}, it is not clear whether
this strategy is more efficient than one in which each semi-coherent stage
analyzes only fresh data. (E.g, the first stage analyzes $20$
days of data and generates candidates, the second semi-coherent stage
searches for  those candidates
in the {\it next} $50$ days of data and generates a list of candidates that
have still ``survived'', these survivors are searched for again in the
{\it next} $150$ days of data, etc.)
In this paper, we investigate both kinds of schemes: schemes where
previously analyzed data is always {\it recycled} into subsequent
stages, and schemes where each semi-coherent stage analyzes only fresh data.

\item[3)] For simplicity, BC ignored the fact that the GWs have
two possible polarizations (in effect pretending that the detectors measure a
scalar wave). This is a reasonable approximation when estimating
the number of grid-points needed to cover the parameter space, but
not, say, when trying to estimate the FA and FD rates
as a function of the threshold at some intermediate stage in the
search. (Roughly speaking, scalar waves with the same matched-filter SNR
would be easier to detect than actual GWs, since with GWs the full
SNR is ``split'' between the two polarizations, in a way that is
unknown {\it a priori}.) In this paper we aim to
make realistic estimates of a GW pulsar's detectability for a
given matched-filter SNR (and given region of parameter space to be
searched over), so we take polarization into account wherever it makes
a significant difference. In practice, this just means that we use
the $\F$-statistic, Eq.~(\ref{eq:fstatdef}), as our detection
statistic.

\item[4)] When estimating computational costs, BC assume that the
demodulations will be done using \emph{stroboscopic
re-sampling}, a method modeled closely on the FFT algorithm.
A different demodulation method, which we shall refer to as the
SFT method, is currently being used by the GW pulsar search codes in
the  LIGO Scientific Collaboration (LSC) software library~\cite{lscsoft}.
The SFT method takes as its input a  short FFT database
(FFT'ed sets of short-time data stretches), and can be more efficient
than \emph{stroboscopic re-sampling} in cases where only a narrow
frequency range of the demodulated time series is of interest.
In this paper we explore the possibility of using different demodulation methods at
different stages of the search, and attempt to find the most efficient
combination.
\end{description}
All the above points will be elaborated on in later sections of the
paper.

\subsection{The general optimization scheme}
\label{subsec:optimize}

In this section we further discuss our search algorithm and its
optimization.  First we establish some notation.  Let $n$ be the total
number of semi-coherent
stages.  Let $N^{\ib}$ be the number of stacks used in the $i^{\th}$
stage and $\Tcoh^{\ib}$ be the length of each stack; the superscript $i$
will always refer to the $i^{\th}$ semi-coherent stage.  The
resolution of the template grid used to cover parameter space is given
in terms of the maximum fractional mismatch in signal power
$\mu^\ib_\max$ \cite{owen}.  Our detection statistic is $\rho^{\ib}$,
the sum of the $\F$ values from the different stacks (obtained after
sliding appropriately):
\be
\rho^{\ib} = \sum_{k=1}^{N^{\ib}}\F_k^{\ib} \,.
\ee
Denote the distribution of $\rho^{\ib}$ in the absence of any signal
by $p(\rho^\ib)$. In the presence of a GW signal of amplitude $h_\RMS$,
let the distribution of $\rho^{\ib}$ at the gridpoint nearest the
actual signal be $p(\rho^\ib | h_\RMS,\mu_\max^{\ib})$.
Let $\rho_\th^\ib$ be the $i^{\th}$-stage threshold, which a
candidate must exceed to advance to the next stage.
that we use to reject candidates or advance them to the next stage.
The $i^{\th}$-stage FA rate  (per candidate) $\alpha^\ib$ and FD rate
(per candidate)
$\beta^\ib$ are given by
\ba
\alpha^\ib(\rho_\th^\ib) &=& \int_{\rho_\th^\ib}^{\infty}p(\rho)d\rho
\,, \label{eq:falsealarm}\\
\beta^\ib (\rho_\th^\ib; h_\RMS ,\mu_\max^{\ib}) &=& \int_0^{\rho_\th^\ib}
p(\rho|h_\RMS, \mu_\max^{\ib} )d\rho \,  .
\label{eq:falsediss}\ea
Practically identical formulae apply to the final, coherent stage as well.

Again, we require of any search algorithm
that, at the very
end of the search, it results in a false detection less
than $1\%$ of the time. Given this constraint, we parametrize the
search's sensitivity by the signal amplitude $h_{th}$ such that
an embedded signal with $h_\RMS > h_{th}$ would be detected
$\agt 85-90\%$ of the time. We enforce the latter condition as follows.
We set the first-stage threshold $h_{\th}$
such that a signal of amplitude
$h_\RMS = h_{\th}$ will pass to the second stage
$90\%$ of the time. At all subsequent stages we set the threshold
such that the same signal with strength $h_\RMS = h_{\th}$ has a
$99\%$ chance of passing to the next higher stage.
That is, we adjust the
the $i^{\th}$-stage threshold $\rho_\th^\ib$
so that $\beta^{(1)} = 0.10$, while $\beta^\ib = 0.01$ for $i > 1$,
and $\beta^{({\coh})} = 0.01$ as well.
The motivation behind making $\beta^{(1)}$ lower than
$\beta^\ib$ for $i > 1$ is the following: We believe that a computationally efficient
algorithm will have the property that a true signal that is strong
enough to pass the first-stage threshold {\it should} generally pass over all the
others. Any source that is not sufficiently strong to make it through to
the end of the detection pipeline should be discarded as soon as possible,
so as not to waste computing power.  This reflects the basic idea
behind our hierarchical searches: to eliminate unpromising regions of parameter
space as quickly as possible, so that computational resources can be
focused on the more promising regions.
Basically, the first-stage threshold determines the
sensitivity of the whole search, and
subsequent steps whittle down the number of candidates
(i.e., the number of small patches in parameter space that perhaps contain a
true signal) until any remaining patches can be fully, coherently analyzed.

To fully specify our search algorithm, we have to choose the
parameters $\Gamma \equiv (n,
\{N^\ib\}, \{\Tcoh^\ib\}, \{\mu_\max^\ib\}, \mu_\max^{coh})$,
where $n$ is the number of semi-coherent stages and
$i = 1,...,n$.
In doing so, we are subject to certain
requirements or constraints:
\begin{itemize}
\item The total amount of data
available is no more that some $T_\max$ (say, 1 year).

\item We wish to detect (with $\sim 90 \%$ FD rate and $< 1 \%$ overall
FA rate) any unknown signal of amplitude $h_\RMS$ greater than
$h_{th}$ (say, $10^{-26}$).

\end{itemize}
Our task is to choose the parameters $\Gamma$ that minimize the
total required computational power $P$, subject to the above constraints.
We arrive at a cost function $P(h_{th})$, the computational cost
of reaching any given sensitivity level. (Really, P is function
of the product $h^2_{th} T_{\rm max}/S_n(f)$, but we are regarding
$T_{\rm max}$ and $S_n(f)$ as fixed.)
We can immediately invert this function to
determine $h_{th}(P)$, the sensitivity achievable for any given
computing power.

Let us first deal with the constraint on the total amount of data.
We are going to consider simultaneously two different modes of all-sky
searches.
In ``data recycling mode,'' at each stage we start back at the beginning of the
data, but take progressively larger values of $N^{(i)}\Tcoh^{(i)}$.  Thus the
first stage looks at data in the interval $[T_0, T_0 + N^{(1)}\Tcoh^{(1)}]$, the
second stage looks at $[T_0, T_0 + N^{(2)}\Tcoh^{(2)}]$ and so on.  The total
observation time is thus
\begin{equation} \label{eq:tobsrecyc}
\Tobs = N^{(n)}\Tcoh^{(n)}\,.
\end{equation}
In ``fresh-data'' mode,
rather than always starting over from the beginning, we analyze fresh
data at each stage. The first stage looks at data in the range $[T_0, T_0 +
N^{(1)}\Tcoh^{(1)}]$, the second stage looks at  $[T_0 + N^{(1)}\Tcoh^{(1)}, T_0 +
N^{(1)}\Tcoh^{(1)} + N^{(2)}\Tcoh^{(2)}]$, etc. The total observation time is thus
\begin{equation}\label{eq:tobsfresh}
\Tobs = \sum_{i=1}^{n}N^{(i)}\Tcoh^{(i)}\,.
\end{equation}
In either data-recycling or fresh-data mode,
one constraint is that $\Tobs \leq T_\max$ where
$T_\max$ is the total amount of data available. Also, in either mode,
at each stage we look only at portions of parameter
space that exceeded the threshold set at the previous stage.

Next we consider our constraints on the overall FA and FD rates
for the pipeline.
The final, coherent follow-up stage is expected to be much more
sensitive than any of the preceding steps; therefore the overall FA rate
is essentially set by the final stage threshold alone.
(The earlier stages serve only
to whittle down the number of candidates, $N_{coh}$, that are
analyzed in the final coherent stage.)
If the threshold in the final follow-up stage is
$\rho_\th^{(coh)}$, then the overall FA rate is no larger than
${\alpha}^{(coh)}(\rho_\th^{(coh)})$ times the number of effectively
independent candidates in parameter space. We approximate the latter, crudely,
by $\sim N_p(T_{\rm max}, 0.2, 1)$;
in practice ${\alpha}^{(coh)}(\rho_\th^{(coh)})$
turns out to be so minuscule that the crudeness of this approximation is
irrelevant.

The overall false dismissal requirement is also easily handled.  Let
$\tilde{\beta}$ be the total false dismissal rate of the multistage
search.  Each stage has its own threshold $\rho_\th^\ib$ and
corresponding false dismissal rate $\beta^\ib$.  If each stage,
including the follow-up stage, were to analyze completely independent
data, we would have
\be
\tilde{\beta} = 1 - \prod_{i=1}^{n+1} (1-\beta^\ib) \approx
\beta^{(1)} + \cdots \beta^{(n+1)} \,.
\ee
(where we use ``$\beta^{(n+1)}$'' interchangeably with $\beta^{(coh)}$).
In our fresh-data search mode, the data at different stages are
independent, {\it except} for the final, follow-up stage.
And in our recycled-data scheme, the
data examined at higher stages includes all the data examined in earlier
stages. Then when $\beta^{(1)}=0.1$ and
\be\label{eq:beta}
\beta^{(2)}= \beta^{(3)} = ... = \beta^{(n)} = \beta^{(coh)} = 0.01 \, ,
\ee
it is clear that $\tilde \beta$ is roughly in the range
$(10+n-1)\% $ to (10 + n)\% $, for fresh-data mode and $10\% $ to
(10 + n)\% $
for recycled-data mode. Since $n \approx 3$ turns out to be optimal
(see below), we crudely summarize this by saying that our
strategies have an overall FD rate of $10-15\%$ at the
threshold value of $h_\RMS$.

Finally, we turn to the search's computational cost, which we wish to
minimize. Let us denote the total number of floating point operations for the
$i^{th}$ semi-coherent stage by $C^\ib$ and for the final coherent stage by
$C^{(\coh)}$.  Expressions for $C^\ib$ and $C^{(\coh)}$ are given in the
next section. For now, it is sufficient to say the total computational
cost is
\be C_{total} = \big(\sum_{i=1}^n C^\ib\big) +
C^{(coh)}\, ,
\ee
and that if we wish to analyze the data in roughly real time,
the required computational power (operations per unit time) is
\begin{equation} \label{eq:comppower}
P = \frac{C_{total}}{\Tobs} \, .
\end{equation}
Depending on which mode we are working in, $\Tobs$ is given by
Eq.~(\ref{eq:tobsrecyc}) or Eq.~(\ref{eq:tobsfresh}).

Again, our strategy for optimizing the search is to minimize $P$,
subject to the constraints listed above.

\section{Template counting, confidence levels, and computational cost}
\label{sec:formulae}

\subsection{Template counting formulae}
\label{subsec:metric}

This section gives the template counting formulae originally derived
by BC using the metric formulation developed in \cite{owen}.

For simplicity, the parameter space is covered by spheres of proper radius
$\sqrt{\mu_\max}$ ($\mu_\max$ is the maximum allowed fractional
mismatch in the detection statistic \cite{owen}) using a cubic grid.
However it is worth keeping in mind that a cubic grid over-estimates the
number of required templates even in two
dimensions, and the difference increases rapidly with the
dimensionality \cite{jhc}.

As mentioned earlier, for each semi-coherent stage,
we have a coarse grid for the demodulation and a fine grid for the
stack-slide analysis.
Following BC, for simplicity we shall require that at any given
semi-coherent stage, the maximal mismatch $\mu_\max$ for the fine
grid is the same as $\mu_\max$ for the coarse one.
However (unlike BC), we allow $\mu_\max$ to vary from one stage to
the next.

The number of templates (or gridpoints) $N_p$ is a function of
the mismatch $\mu_\max$, the coherent time baseline $\Tcoh$, and the
number of stacks $N$ (which is unity for the coarse grid).
BC have derived the following expressions for the number of
gridpoints, $N_{pc}$ and $N_{pf}$, in the coarse and fine grids,
respectively:
\ba
N_{pc} &=& N_p(\Tcoh, \mu_\max, 1)\,, \label{eq:npc}\\
N_{pf} &=& N_p(\Tcoh, \mu_\max, N)\,. \label{eq:npf}
\ea
where $N_p$ is given in Eq.~(2.22) of BC:
\be \label{eq:Np}
N_p = \mathop{\textrm{max}}_{s\in
  \{0,1,2,3\}}\left[\mathcal{M}_s\mathcal{N}_sG_s\prod_{k=0}^s\left(1+
  \frac{0.3r\Omega^{k+1}\tau_\min^k}{c\,k!\sqrt{\mathcal{M}_s}}\right)\right]  \, .
\ee
Here $r=1\,$AU is Earth's orbital radius, $\Omega = 2\pi/(1 {\rm yr})$,
\be
\mathcal{N}_s = \frac{s^{s/2}}{(s+2)^{s/2}}\frac{f_\max^s
  \Tcoh^{s(s+3)/2}}{(\mu_\max/s)^{s/2}\tau_\min^{s(s+1)/2}}\, ,
\ee
\be
\mathcal{M}_s = \big(\frac{f_\max}{1 {\rm Hz}}\big)^2\frac{(s+2)}{4\mu_\max}\left( \frac{1}{A^2} +
\frac{1}{B^2} + \frac{1}{C^2}\right)^{-1/2}\, ,
\ee
where
\be
A = 0.014\ , B = 0.046\left(\frac{\Tcoh}{1\textrm{ day}}\right)^2 \,
C = 0.18\left(\frac{\Tcoh}{1\textrm{ day}}\right)^5 \,
\ee
and the functions $G_s$ are given in Appendix A of BC.
Roughly speaking, the factor $\mathcal{M}_s$ counts distinct patches
on the sky as set by the Earth's one-day spin period, $\mathcal{N}_s$
counts distinct ``patches'' in the space of spin-down parameters,
the $G_s$ give the dependence of $N_p$ on the number of stacks, $N$, and the
factors of the form $\left(1+
  \frac{0.3r\Omega^{k+1}\tau_\min^k}{c\,k!\sqrt{\mathcal{M}_s}}\right)$
effectively account for the increase of search volume required when
the frequency derivative $d^kf/dt^k$ is dominated by the Doppler shift
from the Earth's motion around the Sun rather than by the pulsar's
intrinsic spin-down. In our numerical work we use the full expressions for
the $G_s$ given in the Appendix A of BC, but for completeness we
note that BC also give the following approximate fits to the
$G_s$, which are valid when  $N\gg 4$:
\ba
G_0(N) &= & 1 \,,\\
G_1(N) &\approx & 0.524N \,,\\
G_2(N) &\approx & 0.0708N^3 \,,\\
G_3(N) &\approx & 0.00243N^6 \,.
\ea
The $N_p$ results in BC were derived under the assumption that the
observation time is significantly less than one year.  As
we shall see below, in the cases where the
total available data covers an observation time of a year or more,
it turns out
that for the optimal search, the initial semi-coherent stages
typically analyze a few days' to a few months'
worth of data.
Also, most of the search's computational cost
is spent on these early stages. (This is especially true
for the young-pulsar search, which is the most computationally challenging.)
Therefore, it seems reasonable for our purposes
to simply use the $N_p$ formulae from BC for {\it all} observation times.
Since the cost-errors we make by using the BS formulae
will be confined to the later
stages, and since the overall sensitivity of the search is effectively set
at the first stage, we believe these errors will not significantly
affect the total computational cost, for fixed threshold (though
they may affect the relative allocation of resources between
the different stage).
Of course, the validity of this assumption can only really
be checked by re-doing the calculation
using more accurate expressions for the $N_p$'s,
appropriate for year-long observation times, but unfortunately such
expressions are not currently available.

Even for short observation times, the
$N_p$ calculation in BC used the approximation
(\ref{eq:approxF}), which neglects the amplitude
modulation of the signal; however this approximation is not expected to cause
significant errors in estimating template numbers.

\subsection{False dismissal rates and the thresholds}
\label{subsec:fd}

In this subsection, we discuss the statistical properties of the
stack-slide search and solve the false dismissal constraint to obtain
expressions for the thresholds.

It is shown in \cite{jks} that the distribution of the $\F$-statistic
(or to be more precise, $2\F$), for each coherent search, is given by
a non-central $\chi^2$
distribution.  The non-centrality parameter $\eta$ is given in terms
of the signal $h(t)$ by:
\ba
\eta &=& 4 \left(1-\frac{\mu_\max}{3}\right)
\int_0^\infty \frac{|\tilde{h}(f)|^2}{S_n(f)} df \nonumber \\
&=& \left(1-\frac{\mu_\max}{3}\right)\frac{2 h_\RMS^2\Tcoh}{S_n(f)}\,,
\label{eta}
\ea
where $\tilde{h}(f)$ is the Fourier transform of $h(t)$.
We have included a fitting factor of $1-\mu_\max/3$ to account for
the average loss in power due to the mismatch between the signal and
template.
$h_\RMS$ is the root-mean-square value of the signal $h(t)$.
We can relate $h_\RMS$
to the amplitude $h_\RMS$ defined in Eqs.~(\ref{eq:waveformplus}) and
(\ref{eq:waveformcross}), as follows. If one averages
$h_\RMS$ over all sky-positions as well as over the polarization
parameters $\iota$ and $\psi$, one obtains
$<h^2_\RMS> = (2/25)h^2_0$ (see Eq. (93) of \cite{jks}).

More explicitly, the distribution is
\ba
p(\F|\eta) &=& 2\chi^2(2\F|\eta,4) \nonumber \\ &=&
\left(\frac{2\F}{\eta}\right)^{1/2} I_1(\sqrt{2\F\eta})e^{-\F-\eta/2}
\ea
where $\chi^2(\cdot|\eta, \nu)$ is the $\chi^2$ distribution with
$\nu$ degrees of freedom and non-centrality parameter $\eta$, and
$I_1$ is the modified Bessel function of first order.
The statistic $\rho$ of interest for the stack-slide
search is the sum of the $\F$-statistic over $N$ stacks.
Assuming the $\F$-statistic for the $N$ stacks to be statistically
independent, $2\rho$
must follow a $\chi^2$ distribution with $4N$ degrees of freedom and
non-centrality parameter $N\eta$
\be
p(\rho|\eta,N) = 2\chi^2(2\rho|N\eta,4N)\,.
\ee
The mean and variance of $\rho$ are given respectively by
\be \label{eq:meanvar}
\bar{\rho} = 2N + \frac{N\eta}{2} \,, \qquad \sigma_\rho^2 = 2N + N\eta\,.
\ee
Using the distribution $p(\rho^\ib)$, the false alarm rate
for the $i^{th}$ semi-coherent stage (defined in
Eq.~(\ref{eq:falsealarm})) can be evaluated analytically:
\begin{equation}
\alpha^\ib(\rho_\th^\ib) =
e^{-\rho_\th}\sum_{k=0}^{2N^\ib-1}\frac{(\rho_\th^\ib)^k}{k!}  \,. \label{eq:fa-analytical}
\end{equation}
As discussed earlier, the overall false alarm probability $\tilde{\alpha}$
for the search is set by the final coherent follow-up stage.  For this
stage, $N=1$ so that if the threshold on $\rho$ is $\rho_\th^{(coh)}$,
then it is easy to see from the previous equation that:
\be \label{eq:alphacoh}
\tilde{\alpha} = (1+\rho_\th^{(coh)})e^{-\rho_\th^{(coh)}}\,.
\ee
In the presence of a signal, the non-central $\chi^2$ distribution for
$\rho$ is a little cumbersome to work with, and it is useful to
replace it by a Gaussian with the appropriate mean and variance.  So
we say that the distribution of $\rho$ must be approximately Gaussian
with mean and variance as in eq. (\ref{eq:meanvar}):
\be
p(\rho|\eta,N) = \frac{1}{\sqrt{2\pi\sigma_\rho^2}}
\,\,e^{-(\rho-\bar{\rho})^2/2\sigma_\rho^2} \, .
\ee
This approximation is not valid when $N$ is of order unity.
Then for any given $h_{\th}$, we should set the threshold of the
$i^{th}$ stage, $\rho^{\ib}$ by the false dismissal requirement:
\be\label{rhoi}
\int_{0}^{\rho^{\ib}} p(\rho|\eta^{\ib}_\th,N) d\rho = \beta^\ib \,,
\ee
where
\be
\eta^{\ib}_\th := \left(1-\frac{\mu_\max}{3}\right)\frac{2h_\th^2\Tcoh^{\ib}}{S_n(f)}\,.
\ee
Here the factor of $1-\mu_\max^\ib/3$ accounts for the average
loss in power due to the mismatch between the signal parameters and
nearest gridpoint parameters.
Eq.~(\ref{rhoi}) can be solved to find $\rho^\ib_\th$ as a function
of $h_{\th}$, $\Tcoh^{\ib}$, and $\mu^{\ib}$.
Or equivalently, it gives
$h_{\th} = h_{\th}(\rho^{\ib}, \Tcoh^{\ib}, \mu^{\ib})$.
This equation can easily be solved by using the properties of the
complementary error function.  By changing variables in the integral,
we can rewrite the false dismissal rate as
\be
\beta^\ib =
\frac{1}{2}\textrm{erfc}\left(\frac{\bar{\rho}^\ib-\rho^\ib}{\sqrt{2}\sigma_\rho^\ib}\right)\,.
\ee
If $h_\th$ is the smallest value of $h_\RMS$ for which the false
dismissal rate is no bigger than $\beta^\ib$, then we have
\ba\label{eq:thresholdsolve}
\rho^\ib(h_\th) &=& \bar{\rho}^\ib -
\sqrt{2}\sigma_\rho^\ib\,\textrm{erfc}^{-1}(2\beta^\ib) \nonumber \\
&\approx& 2N^\ib\left[1 + \frac{\eta_\th^\ib}{4}\right] \nonumber \\ &-&
2\textrm{erfc}^{-1}(2\beta^\ib) \sqrt{{N}^\ib}\sqrt{1 +\frac{\eta_\th^\ib}{2}}\,.
\ea

In practice, we fix one value of $h_\th$ (our sensitivity goal)
for an an entire search,
and we then set the threshold $\rho^{\ib}$ at each stage by solving
Eq.~(\ref{eq:thresholdsolve}), with the false dismissal rates set by
$\beta^{(1)} = 0.1$ and $\beta^\ib = \beta^{\rm coh} = 0.01$ for
$i\geq 2$.
Our rationale for this choice is as follows.
At each stage, one can estimate the signal strength of any
successful candidate. If after the first stage, one can already predict
that a candidate is not strong enough to pass over the threshold at the
second or a higher stage, then one might as well discard it immediately
and so not waste computer power on a likely failure.
Put the other way, an efficient algorithm should ensure that
a true signal that is strong enough to pass over the first stage is also
strong enough to pass over all subsequent stages. Then the
false dismissal rate for the whole search will be only a little larger
than the FD rate of the first stage alone, or a little more than $10\%$.
(An {\it overestimate} of the total FD rate is the sum of the rates for
each of the stages, or $13\%$ for a 3-stage search.)

\subsection{Computational Cost}
\label{subsec:cost}

Let us begin with the first semi-coherent stage.  Here, the number of
points in the
coarse and fine grids are respectively
\ba
N^{(1)}_{pc} &=& N_p(\Tcoh^{(1)},\mu_\max^{(1)}, 1) \,, \\
N^{(1)}_{pf} &=& N_p(\Tcoh^{(1)},\mu_\max^{(1)},N^{(1)}) \,.
\ea
If we are searching in a frequency range from small frequencies up to $f_\max$,
the data must be sampled in the time domain (at least) at the Nyquist
frequency $2f_\max$.  The minimum number of data points that we must
start out with in the time domain is then $2f_\max\Tcoh$. To calculate
the $\F$-statistic for each stack, we need to first calculate the
quantities $F_a$ and $F_b$ which appear in equation
(\ref{eq:fstatdef}).  We describe two methods below which
may be called the \emph{stroboscopic resampling method} and the
\emph{SFT method}. Given $F_a$ and $F_b$, the cost of combining
them to get $\F$ is negligible.\\

\noindent \textit{The stroboscopic resampling method:} The method
suggested in \cite{jks} (and also in \cite{bc}) is based on the
observation that the integrals in Eqs.~(\ref{eq:Fa}) and (\ref{eq:Fb})
look \emph{almost} like a Fourier transform; the difference being the
form of $\Phi(t)$ in the exponential.  However, by suitably resampling
the time series, effectively redefining the time variable so that the
spectrum of a real signal would look like a spike in a single frequency
bin, the integral can be written as a Fourier transform and we can
then use the FFT algorithm.  Since the cost of calculating an FFT for
a time series containing $m$ data points is $3m\log_2 m$, the operations
cost of
calculating the $\F$-statistic for each stack should be approximately
$12f_\max \Tcoh\log_2(2f_\max\Tcoh)$.  Repeating this for $N^{(1)}$
stacks and for each point in the coarse grid, we see that the total
cost of calculating $F_a$ and $F_b$, and therefore the $\F$-statistic,
is approximately
\be
12N^{(1)}N^{(1)}_{pc}f_\max\Tcoh^{(1)}\log_2(2f_\max\Tcoh^{(1)})\,.
\ee
We now need to appropriately slide each segment in frequency space and
stack them up, i.e. add the $\F$-statistic values from each
stack to get our final statistic $\rho$.  This has to be done for each
point in the fine grid.  The cost of sliding is negligible and we need
only consider the cost of adding the $\F$-statistic values.
Since adding $N^{(1)}$ real
numbers requires $N^{(1)}-1$ floating point operations, we see that
the cost of stacking and sliding for all frequency bins and for all
points in the fine grid is approximately
\be
f_\max\Tcoh^{(1)}{N_{pf}^{(1)}}
(N^{(1)}-1)  \,.
\ee
Thus, the computational cost for the first semi-coherent stage is
\ba\label{eq:hierarchical-cost-1}
C^{(1)}_\res &=& f_\max\Tcoh^{(1)}N_{pc}^{(1)}\left[
12N^{(1)}\frac{\log(2f_\max\Tcoh^{(1)})}{\log 2} \right. \nonumber \\
&+& \left.
\frac{N_{pf}^{(1)}}{N_{pc}^{(1)}}(N^{(1)}-1)  \right] \,.
\ea
The subscript $\res$ indicates that this result is for the
stroboscopic resampling method.\\

\noindent \textit{The SFT method:} An alternative method is to
use as input not the time series, but rather a bank of short time
baseline Fourier Transforms (SFTs).  This is in
fact the method currently being used in the search codes of the LIGO
Scientific Collaboration \cite{lscsoft}.
Here one first breaks up the data into short segments of length $\Tsft$,
and calculates the Fourier transform of each
segment. (These segments, which are to be combined \emph{coherently}, are
not to be confused with the segments used in the stack-slide algorithm
which are combined incoherently).  $\Tsft$ should be
short enough so that the signal does not drift by more than half a
frequency bin over this time.
Typical values of $\Tsft$ are $1800$s.
The exact method of calculating the
$\F$-statistic from an SFT database is sketched in Appendix A, and
the operations count is also derived there. The result is
(see Eq.~(\ref{eq:sftcostapp})):
\be
\approx 640 N^{(1)}N_{pc}^{(1)}f_\max \frac{(\Tcoh^{(1)})^2}{\Tsft}
\,\,\textrm{Flops}\,.
\ee
Note that the SFT method of calculating the $\F$-statistic is
$\mathcal{O}((\Tcoh^{(1)})^2)$ while for the stroboscopic resampling
method it is
$\mathcal{O}(\Tcoh^{(1)}\log \Tcoh^{(1)})$.

The total cost of stacking and sliding in the first hierarchical stage
using the SFT method is thus:
\ba\label{eq:c1sft}
C^{(1)}_\sft &=& f_\max\Tcoh^{(1)}N_{pc}^{(1)}\left[ \frac{640 N^{(1)}
\Tcoh^{(1)}}{\Tsft} \right. \nonumber \\  &+& \left.
\frac{N_{pf}^{(1)}}{N_{pc}^{(1)}}(N^{(1)}-1) \right]\,.
\ea
When all frequencies are to be searched over,
stroboscopic resampling produces the $\F$-statistic
about an order of magnitude more cheaply than the SFT method, for typical
values of $\Tcoh^{(1)}$.  However when previous stages have narrowed
the search to a small fraction of the whole frequency band (for any given
$\vec\lambda$), the SFT method can be the more efficient one.
We should also mention here that it is possible to start with SFTs and
combine them in such a way as to get a $\mathcal{O}(\Tcoh^{(1)}\log
\Tcoh^{(1)})$ operations count; this is in fact the method used in
\cite{abjk}. However, in this paper, by the ``SFT method'' we always
mean the method described here in Appendix B, with the operation count
given above in Eq.~(\ref{eq:c1sft}).

It also seems likely that the resampling method could be
modified so as to be the most efficient one, even when only
wanted to demodulate a small
frequency band
\be
\Delta f = \textrm{max}\left\{1,\frac{\Tcoh^\ib}{\Tcoh^{(i-1)}}\right\}
\ee
around every selected candidate.  Presumably the first step would be
to heterodyne the data to shift
the relevant frequency range to a neighborhood of zero-frequency.
Then one would filter out frequencies higher than $\Delta f$,
followed by the usual demodulation.
Eq.~(\ref{eq:hierarchical-cost-i}) would then be modified, so that the
new cost of demodulating would be $12N^\ib\Tcoh^\ib\Delta
f\log_2(2\Tcoh^\ib\Delta f)$.  However since the details of this
modified demodulation method have not yet been worked out, we will
not consider it further in this paper.

This completes our analysis of the first stage computational costs for
both methods.  The analysis for the subsequent stages proceeds
similarly; the only difference is
that subsequent stages analyze only those regions of parameter space
that have not been discarded by any of the previous stages.  Assuming
that almost all the candidates are due to noise, the false alarm rate
is a good estimate of the number of candidates produced by any stage.
Let us denote by
$F^{(i)}$ the number of candidates which survive the $i^{th}$
stage.  Since the false alarm rate for the first stage is
$\alpha^{(1)}$, the number of candidates produced by the first
stage is given by
\be
F^{(1)} =
\textrm{max}\left\{1,f_\max\Tcoh^{(1)}N_{pf}^{(1)}\alpha^{(1)}\right\}
\,.
\ee
Note that we will always have at least one candidate which makes it
through to the next stage.
To calculate the cost of a search, we of course must make some
assumptions about the data to be processed. Basically, we are assuming
that the data consists of Gaussian noise plus one detectable source.
(Though we call $F^{(i)}$ the ``$i^{(th)}$-stage false alarm rate'', it
is really the ``false alarm rate or the true-source survival rate, whichever
dominates''. In practice, until the last semi-coherent stage,
the FA rate always dominates.)

To estimate the computational cost for the $i^{th}$
stage, for $i>1$, recall that each of the $F^{(i-1)}$
candidates produced by the $(i-1)^{th}$ stage is in fact a region in
parameter space.  If we assume that the $i^{th}$ stage
further refines
this region, then we see that the number of $i^{th}$-stage coarse grid
points
in this region must be, on average,
$N_{pc}^{\ib}/N_{pf}^{(i-1)}$ (again,
assuming this ratio to be bigger than 1).
Thus, using the stroboscopic resampling method, the number of floating
point operations to calculate the $\F$-statistic in the $i^{th}$ stage
is
\be
F^{(i-1)} \textrm{max}\left\{
1,\frac{N_{pc}^{\ib}}{N_{pf}^{(i-1)}}\right\}
12f_\max \Tcoh^{\ib}N^{\ib}\log_2(2f_\max\Tcoh^{\ib})\,.
\ee
Each candidate produced by the $(i-1)^{th}$ stage occupies a frequency
band $1/\Tcoh^{(i-1)}$, and thus corresponds to $\Tcoh^\ib/\Tcoh^{(i-1)}$
$i^{th}$-stage  frequency bins.  Thus the operations count
for the stacking and sliding is
\ba
&&F^{(i-1)}
\textrm{max}\left\{1,\frac{\Tcoh^\ib}{\Tcoh^{(i-1)}} \right\}
\textrm{max}\left\{1,\frac{N_{pc}^{\ib}}{N_{pf}^{(i-1)}}\right\}
\times \nonumber \\
&& \times \frac{N_{pf}^{\ib}}{N_{pc}^{\ib}} (N^{\ib}-1)
\ea
floating point operations.
Combining these results, we get the computational cost for the
$i^{th}$ stage ($i\geq 2$):
\ba
C^\ib_\res &=&  F^{(i-1)}
\textrm{max}\left\{1,\frac{N_{pc}^{\ib}}{N_{pf}^{(i-1)}}\right\}
\times \nonumber \\ &\times&
\left[ 12N^\ib f_\max\Tcoh^\ib\frac{\log(2f_\max\Tcoh^\ib)}{\log 2}
\right. \label{eq:hierarchical-cost-i}
\\ &+& \left.  \textrm{max}\left\{1,\frac{\Tcoh^\ib}{\Tcoh^{(i-1)}} \right\}
\frac{N_{pf}^{\ib}}{N_{pc}^{\ib}}(N^{\ib}-1)  \right] \,. \nonumber
\ea

If instead one uses the SFT method for calculating the $\F$-statistic,
it is easy to see the operations count is
\ba
C^\ib_\sft &=&  F^{(i-1)}
\textrm{max}\left\{1,\frac{N_{pc}^\ib}{N_{pf}^{(i-1)}} \right\}
\textrm{max}\left\{1,\frac{\Tcoh^\ib}{\Tcoh^{(i-1)}}\right\} \times \nonumber \\
&\times&\left[\frac{640 N^\ib\Tcoh^\ib}{\Tsft} +
\frac{N_{pf}^\ib}{N_{pc}^\ib}(N^\ib -1) \right] \,. \label{eq:hierarchical-cost-i-sft}
\ea

After the $n$ semi-coherent steps, we have the final coherent follow-up stage
where the entire stretch of data of duration $\Tobs$ is used. For this stage,
we analyze $F^{(n)}$ candidates and simply compute the $\F$-statistic without
breaking up the data into any smaller stacks. The cost $C^{(coh)}$ for this
using the resampling method is
\be\label{eq:hierarchical-cost-coh}
C^{({\mathrm coh})}_\res =
F^{(n)}\textrm{max}\left\{1,\frac{N_{p}^{(coh)}}{N_{pf}^{(n)}}
\right\} 12 f_\max\Tobs\frac{\log(2f_\max\Tobs)}{\log 2}
\ee
where $N_{pf}^{coh} \equiv N_p(\Tobs,\mu_{coh},1)$, and
$\mu_{coh}$ is the $\mu_{max}$ of the final, coherent stage.
Using the SFT method, we would have
\be
C^{({\mathrm coh})}_\sft = F^{(n)}\textrm{max}\left\{1,\frac{N_{p}^{(coh)}}{N_{pf}^{(n)}}
\right\} \textrm{max}\left\{1,\frac{\Tobs}{\Tcoh^{(n)}}
\right\}\frac{640\Tobs}{\Tsft}
\ee

So far, all results in this section are valid whether we are working in
fresh-data mode or data-recycling mode.  The following formulae,
for the number of candidates which survive a given
stage, do however depend on which mode we are working in.  If we operate
in fresh-data mode (analyzing fresh data at every
stage--except the last stage, which is a coherent follow-up of all
the searched data), we clearly have (for $i\geq 2$)
\ba\label{Fifresh}
F^{\ib} &=& \alpha^{(i)} \textrm{max}\left\{F^{(i-1)}, 1 \right\}
\textrm{max}\left(1,\frac{N_{pf}^{\ib}}{N_{pf}^{(i-1)}}\right) \nonumber
\\&\times& \textrm{max}\left(1,\frac{\Tcoh^{\ib}}{\Tcoh^{(i-1)}}\right) \, .
\ea
Again, our count assumes that at least
one candidate gets ``promoted'' to the succeeding stage.
We note that Eq.~(\ref{Fifresh}) assumes that
the parameter space resolution improves at every stage of
fresh-data mode (which seems always to be true for our
optimized searches).
We also note that Eq.~(\ref{Fifresh}) is basically
identical to Eq.~(5.2) of BC, but there it is
claimed to be the FA rate for data-recycling mode. That is
not correct, in general, as we discuss further below.

If we are in data-recycling mode (at each step, re-analyzing old data,
while also adding on new data), then the probabilities of
a candidate's randomly surviving the $(i-1)^{th}$ and
$i^{th}$ stages are {\it not} independent, and so Eq.~(\ref{Fifresh})
is no longer valid.
(To see this, consider the limit where only a very tiny bit of data is
added on, and the resolution is kept fixed. Then any candidate that
survives the $(i-1)^{th}$ stage has almost a $100\%$ chance of surviving
the $i^{th}$ stage, even if $\alpha^{(i)}$ is extremely small.)
Indeed, the rhs of Eq.~(\ref{Fifresh}) is clearly a {\it lower bound}
on the $i^{th}$-stage false alarm rate, in data-recycling mode.

We can also place the following {\it upper bound} on
on $F^{\ib}$ for data-recycling mode:
\be
\label{FiBC}
F^{\ib} = f_{\mathrm{max}}\Tcoh ^{\ib}\, N_{pf}^{\ib} \alpha^\ib  \, .
\ee
The rhs of (\ref{FiBC}) is the number of false alarms that would
result if one performed a semi-coherent search of the {\it entire}
parameter space with the given $(N^{(i)}, \Delta T^{(i)}, \mu^{(i)},
\rho^{(i)})$, while the lhs is the false alarms that result from
searching only neighborhoods of the points the survived the
$(i-1)^{th}$ stage. 
Thus for data-recycling mode, we can say that
$F^{\ib}$ is somewhere in the range
\ba
\label{FiBC2}
F^{(i-1)} \biggl(\frac{N_{pf}^{\ib}}{N_{pf}^{(i-1)}}
    \frac{\Tcoh^{\ib}}{\Tcoh^{(i-1)}} &\biggr)& \, \alpha^{(i)} \,
\le  \,  F^{\ib} \nonumber \\
&\le & \, f_{\mathrm{max}}\Tcoh ^{\ib}\, N_{pf}^{\ib}
\alpha^\ib  \; .
\ea
Fortunately, when we calculate the total computational cost of some
optimized search in data-recycling mode, needed to achieve some
given sensitivity $h_{th}$, if we try
plugging in {\it either} the upper or lower bound for
$F^{\ib}$, we find the two final results differ from each other
by $\lesssim 18\%$  for a young palsar ($\tau_{min} =40$ year) and
    $\lesssim 5\%$ for an old one ($\tau_{min} =10^6$ year), which for our purposes is
practically insignificant.  Moreover, the optimized search parameters obtained
when we plug in the upper-limit estimate for $F^{\ib}$ are quite similar
to those we find by plugging in the lower limit instead.
Therefore it is safe for us to choose {\it either} the upper or lower limit
as an estimate  of $F^{\ib}$.  For concreteness, in the rest of this
paper we always estimate $F^{\ib}$ by its upper limit, which slightly
overestimates the computational cost of the search.

With these results in hand, we are now ready to calculate the total
computational cost of the entire search pipeline.  We have a number of
choices to make.  At each stage, we can use either the stroboscopic
resampling method or the SFT method in each stage, and we can work in 
either the data-recycling
mode or fresh-data mode from the second stage onwards.
For convenience, we somewhat arbitrarily limit the choices by
considering only strategies that use either
data-recycling mode in every stage or
fresh-data mode in every stage. As we shall see below,
the efficiencies of these two sorts of searches turn out to be
extremely close anyway. Therefore we strongly suspect that more
general searches (using fresh-data mode in some stages and
data-recycling mode in others) would not give significant improvements.

\section{Results}
\label{sec:results}

\subsection{The optimization method}
\label{subsec:optimizationmethod}

We next describe our numerical
optimization method.   The function we want to minimize, the
computational power of Eq.~(\ref{eq:comppower}), is a
complicated function on a large-dimensional space.  Our chosen method
is a simulated annealing algorithm~\cite{mrrtt, kgv} based on
the downhill simplex method of Nelder and Mead~\cite{nm}.  The
downhill simplex method consists of evaluating the function on the
vertices of a simplex and moving the simplex downhill and shrinking it
until the desired accuracy is reached.  The motion of the simplex
consists of a prescribed set of ``moves'' which could be either an
expansion of the simplex, a reflection around a face, or a
contraction. This method is turned into a simulated
annealing method by adding a random fluctuation to the values of the
function to be minimized, at the points of the
simplex.  The temperature of the random fluctuations is reduced
appropriately, or in other words ``annealed'', until the minimum is
found.

There are no universal choices for the rate of annealing or the
starting point of the simplex; these
depend on the particular problem at hand.  For the results presented
below, we have used a variety of different starting points and
annealing schedules to convince ourselves that the optimization algorithm has
converged and that we have indeed
found the best minimum.
Let us first discuss the starting temperature, whose meaning is as
follows.   If $f$ is the
the function to be minimized, then
the temperature
$\Theta$ parametrizes the amplitude of random
fluctuations $f \rightarrow f + \delta f$ added to
$f$ at the points of the simplex:
\be
\delta f  = -\Theta \log r
\ee
where $0 < r < 1$ is a uniformly distributed random number.
A simplex move is always accepted if it takes the simplex downhill, but an
uphill step may also be accepted due to these random fluctuations.
In our case, we found that a starting temperature of $\Theta \sim
10^6$-$10^9$ gives good convergence; this value is to be
compared to the typical value $\sim 10^{13}$ of the
computational cost near its minimum for most of the results presented
below.  We allow a maximum of $500$ iterations of the
simplex.  If the simplex does not converge within $500$ iterations, we
reduce the temperature by $2-5\%$ and restart the iterations from the
best minimum found up to that point.  These steps are repeated until
the simplex converges.   The starting point of the simplex cannot be
chosen arbitrarily, and for this purpose, it is useful to have a rough
idea of the location of the minimum. This requires some experimenting
with a sufficiently broad range of starting points; this is especially
important when the number of variables is large, as is the case for,
say, a search with $n>3$.  Having found a
suitable starting point for one set of pulsar parameters ($f_\max$ and
$\tau_\min$), it can be reused for nearby pulsar parameter values.
Occasionally, to obtain convergence to a minimum, we have taken
as our starting point the minimum we found for nearby pulsar-parameter
values.

We next describe how we impose the constraint that the total amount of
analyzed data is less than $T_\max$.  One could imagine trying to
do this using the method of Lagrange multipliers.
However this seemed difficult to implement numerically  (for our
highly non-linear function $P$), and we found a simpler approach that
suffices. The function our algorithm minimizes is not the total
computational power $P$ (defined in Eq.~(\ref{eq:comppower})) itself,
but rather 
\be \label{eq:compconstr}
f = P \times \left[1 + S\left(\frac{\Tobs}{T_\max}\right)\right]
\ee
where $S(x)$ is a smooth function such that $S(x) = 0$ for $0 < x < 1$ but
S(x) is rapidly increasing exponential function for $x > 1$.
That is, we impose a very steep penalty for leaving the
constraint surface.
This works well, and indeed we found it useful
to impose some additional (intuitively obvious)
constraints in this way, such as
requiring the $N^\ib$ and $\Delta T^\ib$ to all be positive; we again multiply
$P$ by factor that is unity when the constraint is
satisfied but is very large when the constraint is violated.
This ``trick'' is used to find the {\it location} of the minimum, but
of course the results we report are the values of the function $P$ there,
not $f$.

There is one additional technical detail, namely, that our optimization
method is meant for the case of continuous, real variables, while
our variables $N^\ib$ are strictly integers.
We handle this by rounding off $N^\ib$ to the nearest integer while calculating
the cost function $f$, every time it is called.  The downhill simplex
algorithm still treats $N^\ib$ as a
continuous variable, i.e. we allow arbitrarily small changes to
$N^\ib$ when the simplex is moving downhill, but such changes have no
effect on $f$.  We have also tried an alternative approach where
$N^\ib$ is kept as a continuous variable throughout, and rounded off
only at the very end. We have found that the two approaches yield
consistent results.

Finally,
we cross-checked our results using two different implementations of
the simulated annealing algorithm--those of
\cite{nr} and \cite{gsl}--and found that they gave basically equivalent
results in our case.

\subsection{The number of semi-coherent stages}
\label{subsec:nstages}

The first question we want to answer is: what is the optimum number $n$ of
semi-coherent stages to use in the search?
Relatedly, we want to know the most efficient method to use for
the $\F$-statistic calculation (stroboscopic resampling or SFT method) and
best mode to work in (fresh-data mode or data-recycling mode).
To answer this, we
consider an all-sky search for fast/young GW pulsars, by which we mean
a search that goes up to frequency
$f_\max=1000\,$Hz and that can detect pulsars with spindown ages $\tau_{\min}$
as short as $40\,$yr. We assume the amount of data available is
$T_\max=1\,$yr, and ask what is the computational power required to
detect pulsar signals whose $h_\RMS$ is
or above $h_{th}$, given by:
\be\label{eq:h5e}
\frac{h_{th}^2}{S_n(f)} = 2.5\times 10^{-5}\textrm{sec}^{-1}\,.
\ee
This signal strength corresponds to $\sqrt{\eta}\approx 39.72$ for a
full 1-yr observation time with a perfectly matched template.
(Here and below we are implicitly assuming that $S_n(f)$ hardly varies
over the frequency range of the signal.)
We choose the $i^{th}$-stage FD rates $\beta^{(i)}$ as given
in (and just above) Eq.~(\ref{eq:beta}), which, along with the
detection threshold given by Eq.~(\ref{eq:h5e}), determines the
$i^{th}$-stage thresholds $\rho^{(i)}$.
For simplicity, we set $\mu^{(coh)} = 0.2$.
While this is a restriction that we simply put in by hand
(to slightly reduce the space of search parameters to be optimized), we
believe this choice has very little effect on the overall optimized
strategy because, as we shall see shortly, the follow-up stage
usually accounts for only a tiny fraction of the total computational
cost.\footnote{An exception is a search for old pulsars presented in tables
\ref{tab:Fresh-params-million} and \ref{tab:Fresh-cc-million}, where
the cost for the follow-up stage turns out to be non-negligible. This
example in particular will have to be revisited when better formulas
for $N_p$ are available.}  
Thus we are left with $3n$ parameters to be optimized:
$(\Tcoh^\ib,\mu_\max^\ib, N^\ib)$ for $i = 1,\ldots ,n$, subject
to the constraint that the total amount of data analyzed, $T_{used}$
[given by Eq.~(\ref{eq:tobsrecyc}) or (\ref{eq:tobsfresh})] is less than
$1\,$yr.

\begin{figure}
  \begin{center}
  \includegraphics[width=\columnwidth]{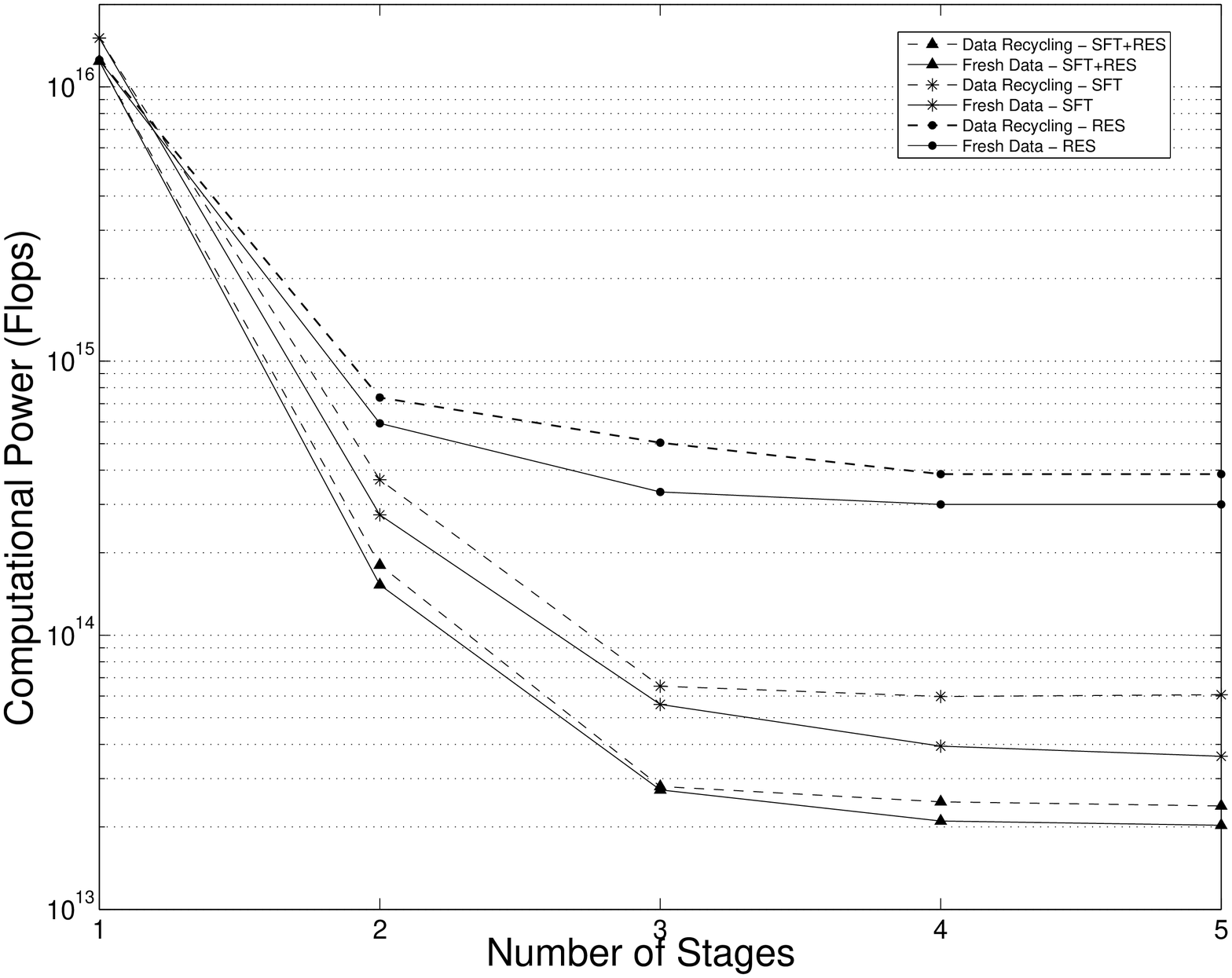}
  \caption{Computational power versus number of semi-coherent
  stages for different methods of calculating the $\F$-statistic. RES
  indicates the stroboscopic resampling method (strategy (ii)) and SFT is
  the SFT method (strategy (i)). SFT+RES corresponds the mixture of these two
  methods (strategy (iii)). For each strategy, solid lines indicate the
  result for the fresh-data mode, while the dashed lines are for the
  data-recycling mode.}
  \label{fig:SFT-RES}
  \end{center}
\end{figure}

Plots of the minimum computational cost for different $n$ and for
both the data-recycling and fresh-data modes are
shown in Fig.~\ref{fig:SFT-RES}.
For each mode, we consider the following three
strategies: (i) Use the SFT method in each stage, (ii) Use the
resampling method in each stage, and (iii) Use the resampling method
in the first and final follow-up stages, and use the SFT method in all
intermediate stages.  Therefore there are $6$ curves in
Fig.~\ref{fig:SFT-RES}.

The most important lessons from Fig.~\ref{fig:SFT-RES} are
the following:  Strategy (iii) turns out to be better than (i) or
(ii).  Furthermore, for strategy (iii), there is a significant
advantage in a three-stage search as compared to a two-stage or single-stage
search, but there is hardly any improvement in computational cost
in going to four or more
semi-coherent stages.
Furthermore, these results are the same
whether we use the fresh-data mode or data-recycling mode, and these
two modes give very similar total costs.
While Fig.~\ref{fig:SFT-RES} presents results just for young/fast pulsar
searches, we find the same basic
pattern for old pulsars, with $\tau_{min} \sim 10^6\,$yr:
strategy (ii) is the most efficient for calculating the $\F$-statistic,
data-recycling mode and fresh-data mode are almost equally efficient, and
having three semi-coherent stages is near-optimal (significantly better than
two stages, and practically as good as four).  The main difference
from the young/fast pulsar case is that the gain in going from 2 to 3
stages is now only a factor $\sim 2$ in computational power, i.e., 
smaller but still significant.    

In the light of these
results, in the rest of this section, we consider only three-stage
searches, with the first stage and final follow-up stages
employing the resampling method and with the second and third stages
employing  the SFT method. We continue to report results for both
data-recycling mode and fresh-data mode.

\subsection{The optimal three-stage search parameters}
\label{subsec:3stage}

For the example search described above, (i.e. $f_\max=1000\,$Hz,
$\tau_{\min}=40\,$yr, $T_\max=1\,$yr, and ${h_{th}^2}/{S_n} = 2.5\times
10^{-5}\textrm{sec}^{-1}$) we list the optimal search
parameters for the three-stage search in data-recycling mode in Table
\ref{tab:bc-params}.  The first two stages analyze about
26 days (divided in 10 segments) and 42 days (divided in 12 segments) of data, respectively,
while the third stage
analyzes the entire year-long data stretch (divided in 8 segments).  The total
computational cost is
$40.2\,$TFlops. The cost breakdown among the individual stages, and
further cost breakdown into the
demodulation piece $C^{(i)}_{coh}$ and the
stack-slide $C^{(i)}_{ss}$ piece in each stage,
are given in Table \ref{tab:bc-cc}.
There we give the total count of floating point operations required, not
the number of operations per second.
\begin{table}[h]
\caption{The optimal search parameters in data recycling
  mode. $f_\max=1000$Hz, $\tau_{\min}=40$yr, $T_\max=1$yr,
  ${h_{th}^2}/{S_n} = 2.5\times 10^{-5}\textrm{sec}^{-1}$, and $\eta$ is
  defined according to Eq. \ref{eta}.}
\vspace{0.2cm}
\begin{tabular}{|c|c|c|c|c|c|}
  \hline
  Stage & $\Tcoh^{(i)}$ (days) & $\mu^{(i)}$ & $N^{(i)}$ & $\Tobs^{(i)}$ (days) & $\sqrt{\eta}$ \\
  \hline
  &&&&&\\
  1 &  2.58 & 0.7805 & 10 &  25.79 &  9.08 \\&&&&&\\
  2 &  3.51 & 0.1139 & 12 &  42.13 &  13.23 \\&&&&&\\
  3 & 45.66 & 0.8196 &  8 & 365.25 &  33.86 \\
  \hline
\end{tabular}
\label{tab:bc-params}
\end{table}
\begin{table}
\caption{The computational cost to analyze one year of  data in
  data-recycling mode. The search parameters are the same as given in
  Table I.  $C^{(i)}_\coh$ is the cost for the coherent
  demodulation step and $C^{(i)}_{ss}$ for the stack-slide step, while
  $C^{(i)}$ is the sum of these two. Follow-up indicates the computational
  cost require for the final follow-up stage.}
\vspace{0.2cm}
\begin{tabular}{|c|c|c|c|}
  \hline
  Stage & $C^{(i)}$ (Flop) &  $C^{(i)}_\coh$ (Flop) & $C^{(i)}_{ss}$ (Flop) \\
  \hline
  &&&\\
  1 & $ 9.37\times 10^{20}$ & $ 6.21\times 10^{19}$ & $8.75\times 10^{20}$  \\
  &&&\\
  2 & $ 3.16\times 10^{20}$ & $ 2.46\times 10^{20}$ & $6.98\times 10^{19} $ \\
  &&&\\
  3 & $ 1.65\times 10^{19}$ & $ 2.73\times 10^{18}$ & $1.37\times 10^{19}$ \\
  &&&\\
  Follow-up &$ 6.30\times 10^{15}$ & & \\
\hline
\end{tabular}
\label{tab:bc-cc}
\end{table}

Our results for fresh-data mode are qualitatively similar, and are
given in Table \ref{tab:fresh-params}.
In this case, the optimal search analyzes about 24 days of data
in the first stage (broken up into 9 segments) and 24 more days in
the second stage (broken into 6 sements).
The third stage analyzes the rest of the year's worth of data, divided into
7 segments. The total computational
requirement is $34.6\,$TFlops and its breakdown is given in Table
\ref{tab:fresh-cc}.

\begin{table}
\caption{Same as Table I, but for fresh-data mode.}
\vspace{0.2cm}
\begin{tabular}{|c|c|c|c|c|c|}
    \hline
Stage & $\Tcoh^{(i)}$ (days) & $\mu^{(i)}$ & $N^{(i)}$ & $\Tobs^{(i)}$ (days) & $\sqrt{\eta}$ \\
    \hline
&&&&&\\
1 &  2.71 & 0.7829 & 9 &  24.35 &  8.82 \\&&&&&\\
2 &  4.08 & 0.0654 & 6 &  24.49 & 10.17 \\&&&&&\\
3 & 45.20 & 0.8229 & 7 & 316.42 & 31.50 \\
\hline
\end{tabular}
\label{tab:fresh-params}
\end{table}
\begin{table}
\caption{Same as Table II, but for fresh-data mode.}
\vspace{0.2cm}
\begin{tabular}{|c|c|c|c|}
    \hline
Stage & $C^{(i)}$ (Flop) & $C^{(i)}_\coh$ (Flop) & $C^{(i)}_{ss}$ (Flop) \\
    \hline
&&&\\
1 & $8.11\times 10^{20}$ & $6.42\times 10^{19}$ & $7.46\times 10^{20}$  \\
&&&\\
2 & $2.64\times 10^{20}$ & $2.62\times 10^{20}$ & $2.67\times 10^{18}$  \\
&&&\\
3 & $1.74\times 10^{19}$ & $5.54\times 10^{18}$ & $1.19\times 10^{19}$ \\
&&&\\
Follow-up & $1.62\times 10^{16}$ & & \\
\hline
\end{tabular}
\label{tab:fresh-cc}
\end{table}

We note the following features of these results.  First, in both modes,
basically all the data has been analyzed by the end of the third
semi-coherent stage.
This is not a requirement that we put in by hand, but rather it arises from the
optimization: the optimal scheme ``gets through'' the entire year's worth of
data before the final follow-up stage.
Secondly, in data-recycling mode, $73.8\%$ of the computing time
is spent in the first stage, $24.9\%$ in
the second, $1.3\%$ in the third and a negligible fraction in the follow-up.  The
results are similar for the fresh-data mode: approximately $74.2\%$ of the
computational resources are spent in the first stage,  $24.2\%$ in the second
stage, $1.6\%$ in the third stage and a negligible amount in the follow up stage.
Finally, fresh-data mode entails
a slightly lower computational cost than data-recycling mode.
However this last fact could be an artifact either of having slightly
different overall FD rates in the two cases, or of our using an overestimate
of $F^{(i)}$ in the latter case. The bottom line is that, after optimization,
the two modes are almost equally efficient.

If instead we consider a search for older pulsars, with $\tau_\min
= 10^6\,$yr instead of $40\,$yr, then the optimal solutionfor both modes  
are summarized in Tables
\ref{tab:BC-params-million}--\ref{tab:Fresh-cc-million}.   A larger
value of $\tau_\min$ means a smaller number of templates, and therefore
a more sensitive search for fixed computational cost.

For data-recycling mode, we have lowered the threshold $h_{th}$ by a
factor $2.35$, to the point where the required computational power is
again $40.2\,$ Tflops, as in the example of Tables \ref{tab:bc-params}
and \ref{tab:bc-cc}. The results are shown in tables
\ref{tab:BC-params-million} and \ref{tab:BC-cc-million}.  
Compared to the young-pulsar search, the computational power is now
spread more evenly over the first two stages: the first stage
consumes about $58.27\%$ of the power, the second stage $38.73\%$,
third stage $3.0\%$ and negligible for the follow-up stage.
\begin{table}
\caption{Search parameters for data-recycling mode with $f_\max=1000$Hz,
  $\tau_{\min}=10^6$yr, $T_\max=1$yr, ${h_{th}^2}/{S_n} = 4.53\times
  10^{-6}\textrm{sec}^{-1}$, and computational power $40.2$Tflops;
$\eta$ is defined according to Eq.~(\ref{eta}).}
\vspace{0.2cm}
\begin{tabular}{|c|c|c|c|c|c|}
    \hline
Stage & $\Tcoh^{(i)}$ (days) & $\mu^{(i)}$ & $N^{(i)}$ & $\Tobs^{(i)}$ (days) & $\sqrt{\eta}$ \\
    \hline
&&&&&\\
1 & 14.84 & 0.3514 & 8 & 118.72 &  9.06 \\&&&&&\\
2 & 30.06 & 0.0917 & 6 & 180.34 & 11.70 \\&&&&&\\
3 & 52.18 & 0.0986 & 7 & 365.25 & 16.63 \\
\hline
\end{tabular}
\label{tab:BC-params-million}
\end{table}
\begin{table}
\caption{The computational cost to analyze one year of  data in
  data-recycling mode. The search parameters are the same as given in
  Table V.  $C^{(i)}_\coh$ is the cost for the coherent
  demodulation step and $C^{(i)}_{ss}$ for the stack-slide step, while
  $C^{(i)}$ is the sum of these two. Follow-up indicates the computational
  cost require for the final follow-up stage.}
\vspace{0.2cm}
\begin{tabular}{|c|c|c|c|}
    \hline
Stage & $C^{(i)}$ (Flop) & $C^{(i)}_\coh$ (Flop) & $C^{(i)}_{ss}$ (Flop) \\
    \hline
&&&\\
1 & $7.41\times 10^{20}$ & $2.85\times 10^{18}$ & $7.39\times 10^{20}$  \\
&&&\\
2 & $4.93\times 10^{20}$ & $3.77\times 10^{20}$ & $1.16\times 10^{20}$  \\
&&&\\
3 & $3.82\times 10^{19}$ & $1.34\times 10^{19}$ & $2.48\times 10^{19}$ \\
&&&\\
Follow-up & $6.18\times 10^{13}$ & & \\
\hline
\end{tabular}
\label{tab:BC-cc-million}
\end{table}

For the case of fresh-data mode, we have lowered the threshold $h_{th}$ by a factor
$2.36$, to the point where the required computational power is
again $34.6\,$ Tflops, as in the example of Tables
\ref{tab:fresh-params} and \ref{tab:fresh-cc}.
Once again, compared to the young-pulsar search, the computational
costs are spread more evenly over the first two stages: the first stage
consumes about $31.3\%$ of the power, the second stage $30.4\%$,
third stage $14.0\%$.  In this case, the cost for the
follow-up stage is $24.0\%$, which is not negligible.  
This indicates that, for this case, the earlier stages have not succeeded in 
reducing the number of candidates to a low level.  
The overall sensitivity, though, is still almost identical to the
data-recycling case.
\begin{table}
\caption{Search parameters for fresh-data mode with $f_\max=1000$Hz,
  $\tau_{\min}=10^6$yr, $T_\max=1$yr, ${h_{th}^2}/{S_n} = 4.47\times
  10^{-6}\textrm{sec}^{-1}$, and computational power $34.6$Tflops;
$\eta$ is defined according to Eq. \ref{eta}.}
\vspace{0.2cm}
\begin{tabular}{|c|c|c|c|c|c|}
    \hline
Stage & $\Tcoh^{(i)}$ (days) & $\mu^{(i)}$ & $N^{(i)}$ & $\Tobs^{(i)}$ (days) & $\sqrt{\eta}$ \\
    \hline
&&&&&\\
1 & 11.77 & 0.2074 & 9 & 105.96 &  9.04 \\&&&&&\\
2 & 10.97 & 0.0199 & 6 &  65.82 &  7.13 \\&&&&&\\
3 & 27.64 & 0.0206 & 7 & 193.47 & 12.22 \\
\hline
\end{tabular}
\label{tab:Fresh-params-million}
\end{table}
\begin{table}
\caption{Same as Table VI, except for fresh-data mode. The 
search parameters are those of Table VII.}
\vspace{0.2cm}
\begin{tabular}{|c|c|c|c|}
    \hline
Stage & $C^{(i)}$ (Flop) & $C^{(i)}_\coh$ (Flop) & $C^{(i)}_{ss}$ (Flop) \\
    \hline
&&&\\
1 & $3.46\times 10^{20}$ & $3.07\times 10^{18}$ & $3.43\times 10^{20}$  \\
&&&\\
2 & $3.35\times 10^{20}$ & $3.33\times 10^{20}$ & $1.91\times 10^{18}$  \\
&&&\\
3 & $1.54\times 10^{20}$ & $9.23\times 10^{19}$ & $6.19\times 10^{19}$ \\
&&&\\
Follow-up & $2.67\times 10^{20}$ & & \\
\hline
\end{tabular}
\label{tab:Fresh-cc-million}
\end{table}

Let us now discuss the false alarm rate.  We require
that the overall FA rate be less than $1\%$, and we
claimed in section~\ref{subsec:optimize} that this is automatically
satisfied in typical, realistic cases.  We can now verify this claim. 
For the $\tau_\min=40\,$yr search summarized in tables
\ref{tab:bc-params}--\ref{tab:fresh-cc}, 
using (\ref{eta}), with $\mu_\max = \mu_\max^{(coh)} = 0.2$, the
threshold corresponds to
$\rho^{(coh)}_\th \approx \bar{\rho} = 2 + \eta/2 \approx 738 $.
By Eq.~(\ref{eq:alphacoh}), this corresponds to
$\tilde{\alpha} \approx 10^{-318}$ (for either mode).  Using
Eq.~(\ref{eq:Np}), the number of independent templates
required for a full coherent search
of the entire parameter space, using $1\,$yr of data, is
$f_\max T_\max N_p(1\textrm{yr},0.2,1) \approx  10^{34}$.
The overall false alarm
rate is thus $FA = f_\max T_\max N_p\tilde{\alpha} \approx 10^{-284} \ll 1\%$
\footnote{Strictly speaking, Eq.~(\ref{eq:Np}) for the number
  of templates is valid only for observation times which are
  significantly less than a year.  However, unless the discrepancy is
  many orders of magnitude, the numbers obtained show that it is
  obviously sufficient for the purposes of this argument. }.

Similarly, for the case $\tau_\min = 10^6\,$yr, $h^2_{th}/S_n = 4.53
\times 10^{-6}$ (data recycling mode, tables
\ref{tab:BC-params-million} and \ref{tab:BC-cc-million})), we get
$\eta \approx 286$ so that $\tilde\alpha \approx 10^{-61}$.  In this
case, have $N_p 
\approx 10^{17}$ so that $FA = f_\max T_\max N_p \tilde{\alpha}
\approx 10^{-34}$. If we look at fresh data mode (tables
\ref{tab:Fresh-params-million} and \ref{tab:Fresh-cc-million}) with
$\tau_\min =  10^6\,$yr, $h^2_{th}/S_n = 4.47 \times 10^{-6}$, we get 
$FA \approx 10^{-33}$. These values are greater than for the case of
young pulsars, but still vastly smaller than $1\%$.   

The basic point is simply this: For an all-sky search, sensitivity is
limited by computing power, so the detection threshold $h_{th}$ in
practice is substantially higher than what it would be for infinite
computing power.  This means that for a signal to be detectable, it
must have quite a high SNR (in the matched filter sense)--which means
that the FA rate is exponentially small.
Being computationally limited means that when we {\it do} detect something,
we can be very confident that it is not simply random noise masquerading as
a signal~\footnote{Of course, no sensible person would ever claim that
he or she had detected a GW pulsar with FA probability of less than
$10^{-284}$. In such a case, the ``statistical error'' is so ridiculously
small that the true FA rate is dominated by the other,
hard-to-quantify factors,
such as the probability of having some bug in the instrumentation or
in the data analysis code.}

\begin{figure}
  \begin{center}
  \includegraphics[width=\columnwidth]{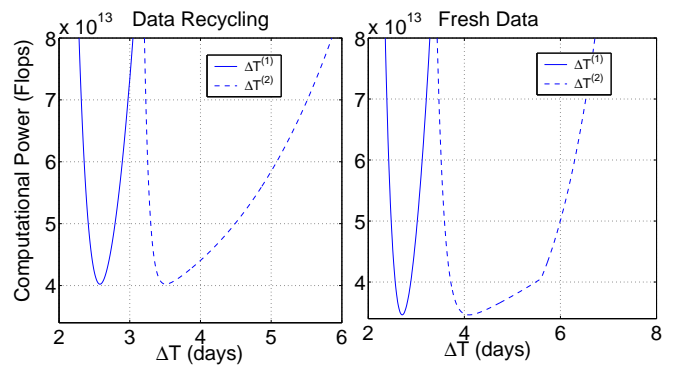}
  \caption{Computational power $P$ as a function of
  $\Delta T^{(1)}$ and $\Delta T^{(2)}$, with all other parameters fixed to their optimal
  values. }\label{fig:DT1}
  \end{center}
\end{figure}

\begin{figure}
  \begin{center}
  \includegraphics[width=\columnwidth]{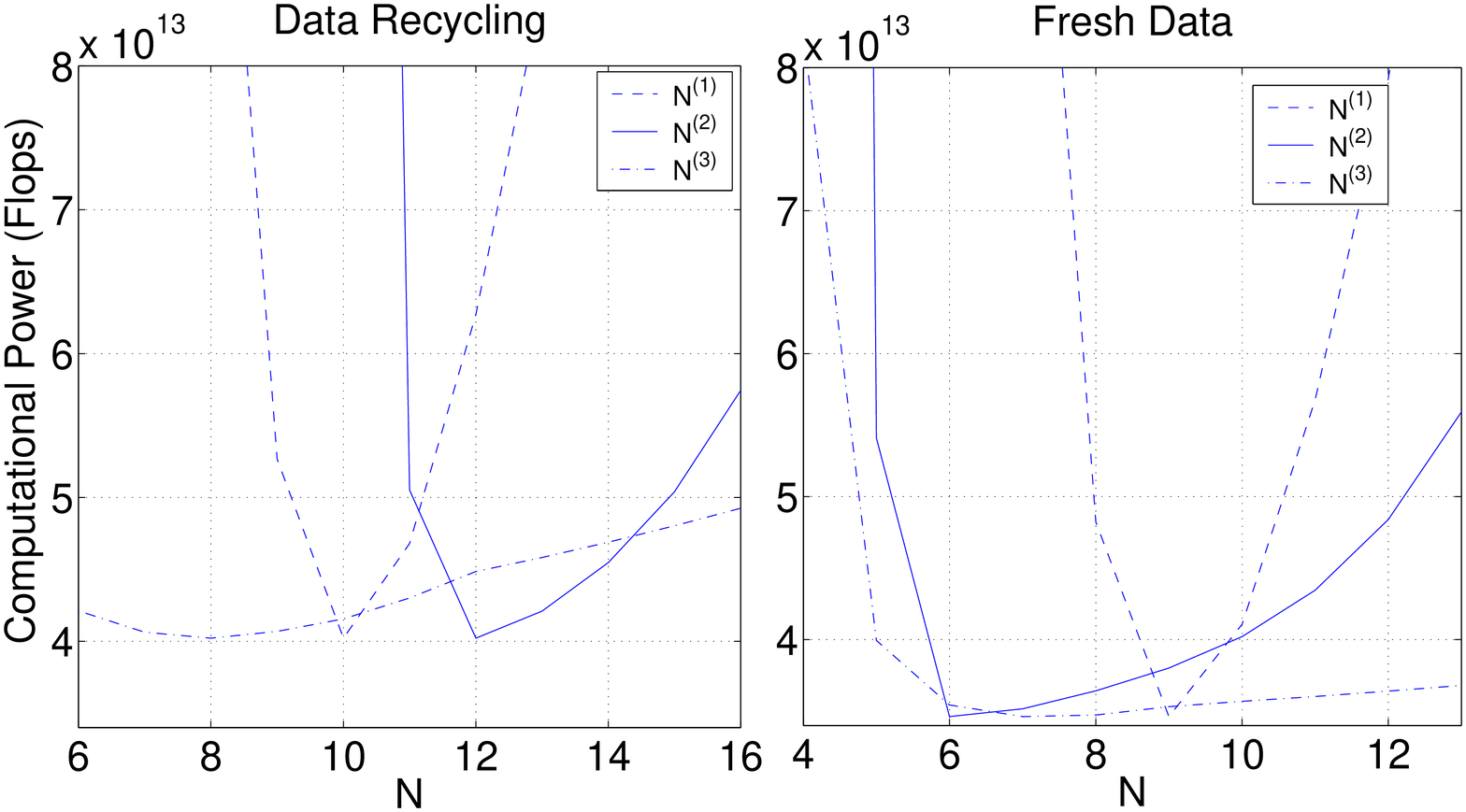}
  \caption{Computational power $P$ as a function of
  $N^{(1)}$, $N^{(2)}$ and $N^{(3)}$.
  For the $N^{(1)}$ and $N^{(2)}$ plots, all other parameters have fixed
  to their optimal values. For the $N^{(3)}$ plot, we have also
  varied $\Tcoh^{(3)}$
  according to $\Tcoh^{(3)} = T_\max/N^{(3)}$ in order to satisfy the
  constraint that the amount of data available is
  $T_\max$.}\label{fig:N1}
  \end{center}
\end{figure}

\begin{figure}
  \begin{center}
  \includegraphics[width=\columnwidth]{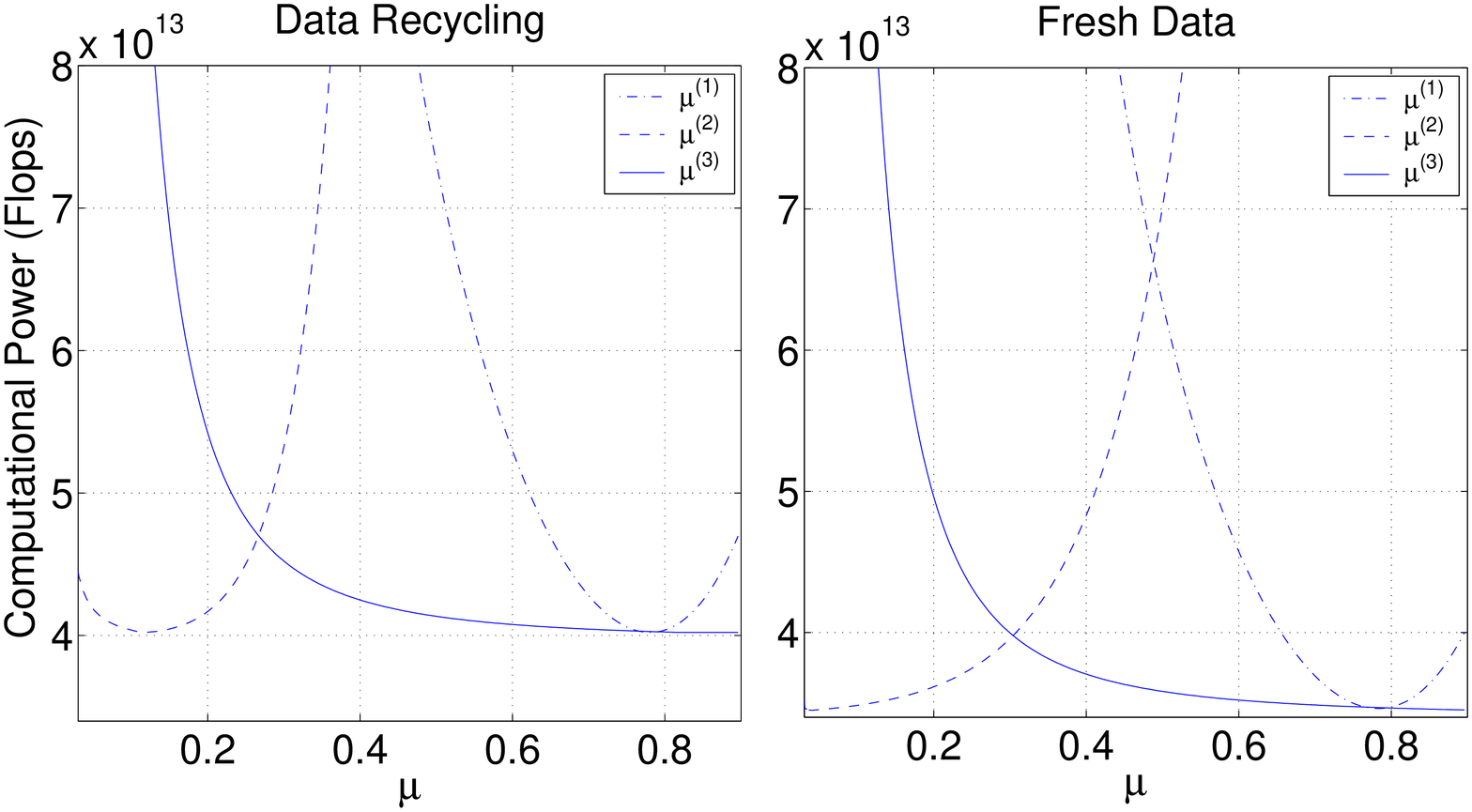}
  \caption{Computational power $P$ as a function of
  $\mu_\max^{(1)}$, $\mu_\max^{(2)}$, and $\mu_\max^{(3)}$.  For each
  plot, all other parameters are fixed at their optimal
  values.}\label{fig:mu123}
  \end{center}
\end{figure}

How accurate are our numerical results?  The total computational cost is a
complicated function on a 9-dimensional space and thus is not easy to
visualize.  We can, however, take appropriate sections of this function
to examine its behavior near the minimum.  Thus, we can ask
whether variations in, say, $\Tcoh^{(1)}$ or $\Tcoh^{(2)}$ away from
their optimal values, increase the computational cost (as they should,
if should if we have truly found a minimum).
To answer this, in Fig.~\ref{fig:DT1} we plot the total
computational power as a function of
$\Tcoh^{(1)}$ and $\Tcoh^{(2)}$, respectively,
for the young-pulsar searches summarized in Tables I-IV.
All the other parameters fixed at their optimal values.
The minima of these curves agree precisely with our simulated
annealing results.
Similarly, Figs.~\ref{fig:N1} and
\ref{fig:mu123} carry the same message, as well as showing
the strong dependence of the computational cost
on $N^\ib$ and $\mu_\max^\ib$.  (It is not so clear from the
plot of $P$ vs.  $\mu_\max^{(3)}$ that this curve has
a minimum in the range shown, but it does in fact have a very shallow
one.)

For the plot of computational cost
versus $N^{(3)}$, we are not allowed to keep $\Tcoh^{(3)}$ fixed, since
that could violate the constraint $\Tcoh^{(3)}N^{(3)} \leq T_\max$.
(Recall that $\Tcoh^{(3)}N^{(3)} = T_\max =1\,$yr for the optimal 3-stage
solution, which is therefore just at the boundary of the constraint
region.)  Therefore, we choose to plot the computational cost as a
function of $N^{(3)}$ while simultaneously varying $\Tcoh^{(3)}$
according to $\Tcoh^{(3)} = T_\max/N^{(3)}$.

A noteworthy feature of these plots is that the
computational power $P$ depends more sensitively on the
early-stage parameters than the late-stage ones; e.g., more
sensitively on $N^{(1)}$ and $N^{(2)}$ than on $N^{(3)}$.
This result should not be surprising since,
as mentioned earlier, for the young-pulsar search the
computational cost of the higher stages is relatively small.

\subsection{The spindown-age and the SNR}
\label{subsec:taumin}

\begin{figure}
  \begin{center}
  \includegraphics[width=\columnwidth]{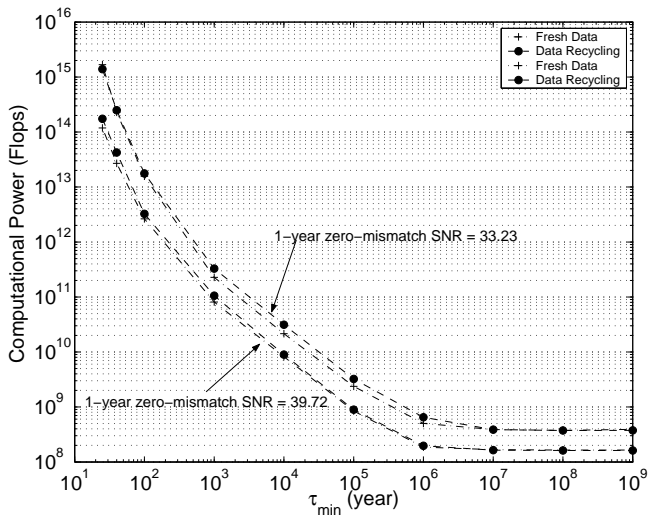}
  \caption{The minimum computational power $P$ required for
  analyzing 1 year's worth of data as a function of the pulsar's
spindown age $\tau_\min$. We
  consider a three-stage search in the both data-recycling and
  fresh-data mode, for two
  different signal strengths. The data-recycling mode results are shown
  with dashed lines, while the fresh-data results are in dotted lines.
  In parts of the curves, the results for the two modes are
  so close together that it is hard to distinguish them.}
\label{fig:bcfresh-taumin}
  \end{center}
\end{figure}

How does the (minimum) computational cost depend on
the shortest spindown timescale that we search over, $\tau_\min$?
Consider again
the case where we have one year of data and we perform an
all-sky search up to a
frequency of $f_\max = 1000\,$Hz.  Fig.~\ref{fig:bcfresh-taumin} shows the
result for the both data-recycling and fresh-data mode,
for two different values
of the 1-year SNR.  Note that these results do pass simple
sanity checks: the computational cost decreases as the SNR
increases (since it is easier to look for stronger signals), and
the computational cost decreases as $\tau_\min$ increases (since it
is easier to search through a smaller parameter space).

\begin{figure}
  \begin{center}
  \includegraphics[width=\columnwidth]{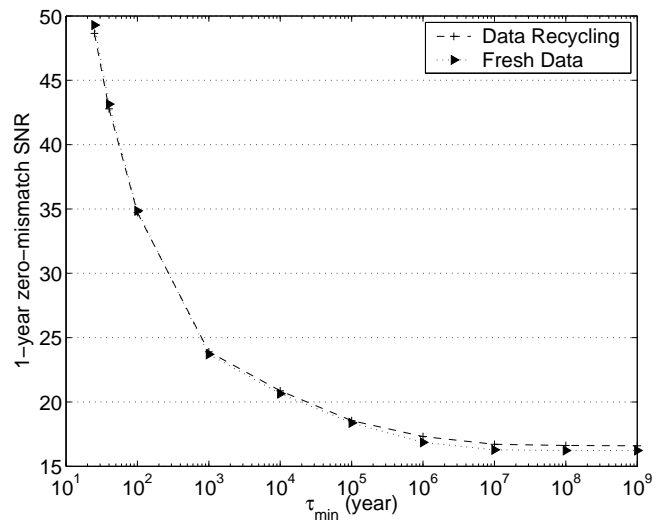}
  \caption{The 1-year SNR (with zero mismatch) as a function
  of  $\tau_\min$, for fixed computational power
  $P= 10^{13}\,$Flops. The dashed line indicates the result for
  data-recycling mode and the dotted line for fresh-data. Since these two
  result are very close to each other, it may be difficult to
  distinguish them.}\label{fig:bcfresh-snr}
  \end{center}
\end{figure}
One can also ask: for a given available computational power, how does
the threshold SNR scale with $\tau_\min$?  This is shown in
Fig.~\ref{fig:bcfresh-snr}. The plot is based on the assumption that
we have one year's worth of data and that we have 10 TFlops of
computing power at our disposal. By ``SNR'', here we mean the
matched-filter SNR, for a perfectly matched filter.
Fig.~\ref{fig:bcfresh-snr} tells us
that a search for unknown GW pulsars with spindown ages
$> 10^6\,$yr  can detect $\sim 85-90\%$ of pulsars whose SNR is $>17$
(again, with FA rate $<< 1\%$).
In an all-sky search for very young pulsars, with $\tau_{min} =
40\,$yr, the SNR required for detection (with the same FD and FA
rates) increases to $\sim 43$.  In comparison, for a source where the
sky position and frequency are known in advance (from radio
observations), an SNR of only $4.56$ is required for detection, with a
$10\%$ FD rate and $1\%$ FA rate~\cite{S1:pulsar}.

Fig.~\ref{fig:bcfresh-snr} strongly suggests that one would
like to simultaneously perform at least two different
all-sky searches: one for old GW pulsars and another for young ones,
with comparable (within a factor ten) computer power devoted to each, but
with quite different thresholds.
(If one set the same threshold for both old and young pulsars, then
almost all computing resources would end up being spent on the young
ones.)
Clearly, to determine the ``best'' apportionment of resources between the
two types of searches would require some additional inputs/assumptions, but
at least Fig.~\ref{fig:bcfresh-snr} seems a good first step towards
making an intelligent allocation.

\section{Conclusions}
\label{sec:conclusions}

Let us first summarize the main results of this paper.  We have
studied general hierarchical strategies for searching for
gravitational waves from unknown, isolated GW pulsars.  In particular, we
have considered multistage hierarchical algorithms where each stage
(except the last)
consists of a coherent demodulation of short stacks of data followed
by appropriate sliding and stacking of the $\F$-statistics results from
the different stacks.
The successive stages serve to quickly reduce the
number of candidates; they are followed by final coherent follow-up stage to
fully analyze the remaining candidates.

We have optimized this strategy by minimizing the computational cost $P$
subject to the constraints which specify the total amount of data
available and the desired confidence levels.
Of course, $P$ depends on the size of the parameter space--
in particular on the range of frequencies and spindown ages that
are searched over.
Carrying out the optimization, and varying over the number $n$ of
semi-coherent stages, we found that the advantages of the multistage
approach saturate at $n=3$ (i.e., $n=4$ and $5$ are scarcely better).

The optimized search parameters $(N^\ib,\Delta T^\ib,\mu^\ib)$
we report should only be considered a rough guide for carrying
out a search in practice because i) in many places we have
used theoretical estimates
of the operations counts instead of those obtained by profiling
existing codes, ii) we have not considered
issues of memory storage or the cost of performing any Monte-Carlo simulations,
and iii) the detector noise has throughout been assumed to be Gaussian and
stationary.  Furthermore, the template counting formulae (\ref{eq:Np})
used in this paper are, strictly speaking, valid only for observation times
significantly less than a year.  The numbers presented in this work
must be recalculated when better approximations become available.
In spite of these limitations, we believe that our results do
provide a useful qualitative guide to what an optimized all-sky search
``looks like.''  In order to optimize actual search codes, applied to
actual data, one must 1) profile the codes to determine the actual
computational cost of the different operations, and 2) do Monte Carlo
studies to determine the {\it actual} $\alpha^\ib$ for different thresholds.
(Recall that the formulae given  here are
based on the assumption stationary, Gaussian noise.) The latter could
require considerable work, especially if the results are strongly
frequency-dependent, with some bands being much ``better behaved''
than others.

Finally, we mention some other possibilities for future work:
\begin{itemize}
\item It would be trivial to extend our work to consider searches that are
  less computationally challenging than all-sky ones, but that are still computationally limited.
  E.g., one could consider searches for unknown NSs in supernova remnants
  (such as SN 1987A), in which case the sky-position is well known but
  the frequency and spindown parameters must be searched over. Similarly, one
  could consider a search over a small fraction of sky, e.g., a portion containing
  the Galactic center or the disk.
\item The formulae for operations counts, confidence levels etc. can
  also be derived for case when the Hough transform~\cite{hough}
  is used in the semi-coherent stages instead of the stack-slide method;
   the optimization of multistage, hierarchical Hough-type searches would
then proceed in the same way as developed here.
\item We expect that the lessons learned in this paper will carry over to
  searches for GW pulsars in low-mass X-ray binaries, which
  are also a computationaly limited~\cite{vecchio}.
  However the details are yet to be worked out.
\item The problem of searching, in LISA data, for the inspiral signals
  of stellar-mass compact objects captured by $\sim 10^6 M_{\odot}$ BHs
  in galactic nuclei, is similar to the GW pulsar search problem, but
  even more computationally challenging~\cite{emri}.
  We expect that the lessons learned in this
  paper will also be very useful in formulating and optimizing a
  search algorithm for LISA capture sources.
\item

In this paper we have tacitly assumed that the search is performed
by a single computer or computing cluster.
However, at least in the next few years,
the most computationally intensive GW searches will be directed by
\texttt{Einstein@Home}\cite{eah}, which relies on tens of thousands
of individual participants donating their idle computing power.
In this case, there might be additional constraints that we
have not yet considered, relating to the rate at which data
and intermediate results can be sent back and forth between the
\texttt{Einstein@Home}\cite{eah} servers and participants' computers, how
much storage is available for use on participants' computers, etc.
We intend to study hierarchical searches in this context also,
to see which if any of the lessons learned here must be modified for the
\texttt{Einstein@Home} context.

\end{itemize}

\section*{Acknowledgements}

The authors thank Bruce Allen, Teviet Creighton, Maria Alessandra
Papa, Reinhard Prix, Bernard Schutz, Xavier Siemens and Alicia Sintes
for valuable discussions. BK acknowledges the hospitality of the
University of the Balearic Islands in Spain.

\begin{appendix}

\section{Computational cost of the SFT method}
\label{app:sftmethod}
Here we estimate the computational cost (in floating point operations)
of calculating the $\F$-statistic via the SFT method.
This result is used in Sec.IV.C.

We begin by reviewing the details of the SFT method; our description
closely follows that given in the documentation of the software
package LIGO Algorithms
Library~\cite{lscsoft}. Imagine
that we wish to compute the $\F$-statistic for a data stretch of length
$\Tcoh$.  Divide this data into
M shorter segments of length $\Tsft = \Delta T/M$,
each containing $N$ data points (so there are
$MN$ data points within $\Tcoh$). The sampled values of
$x(t)$ can then be written as $x_{\alpha j}$ where $0\leq \alpha < M$
labels the segment and $0\leq j < N$ labels points within a
segment. Eq.~(\ref{eq:Fa}) can then be discretized as follows:
\be
F_a(\bmath{\lambda}) = \sum_{\alpha = 0}^{M-1}\sum_{j=0}^{N-1} a_{\alpha j}x_{\alpha
j} e^{-i\Phi_{\alpha j}(\bmaths{\lambda})}\,
\ee
and similarly for Eq.~(\ref{eq:Fb}).
Let $\tilde{x}_{\alpha k}$ be the discrete Fourier transform of
$x_{\alpha j}$ along the index $j$, so that
\be
x_{\alpha j} = \frac{1}{N}\sum_{k=0}^{N-1}\tilde{x}_{\alpha k}e^{2\pi ijk/N}\,.
\ee
Then if we approximate the amplitude
modulation function $a(t)$ as constant over the short-time baseline $\Tsft$,
the expression for $F_a$ becomes
\be
F_a(\bmath{\lambda}) = \sum_{\alpha = 0}^{M-1}a_\alpha
\sum_{k=0}^{N-1}\tilde{x}_{\alpha k} \left[
\frac{1}{N}\sum_{j=0}^{N-1}\exp\left(\frac{2\pi i jk}{N} -
i\Phi_{\alpha j}(\bmath{\lambda})\right) \right] \,.
\ee
The short-time baseline $\Tsft$ is generally chosen so that neither
pulsar spindown nor the Doppler effect causes the signal power to
shift by more than, say, half a frequency bin.  Then we can find
functions $A_{\alpha}(\vec{\lambda})$ and $B_{\alpha
  k}(\bmath{\lambda})$ such that to a good approximation
\be
\Phi_{\alpha j}(\vec{\lambda}) - \frac{2\pi jk}{N}  = A_{\alpha}(\vec{\lambda}) +
\frac{B_{\alpha k}(\bmath{\lambda}) j}{N} \,.
\ee
Thus we have
\ba
&&\frac{1}{N}\sum_{j=0}^{N-1}\exp\left(\frac{2\pi i jk}{N} -
i\Phi_{\alpha j}(\bmath{\lambda})\right) \nonumber \\&& =
e^{-iA_{\alpha}(\vec{\lambda})} \left( \frac{1-e^{-iB_{\alpha
k}(\bmaths{\lambda})}}{1-e^{-iB_{\alpha k}(\bmaths{\lambda})/N}} \right) \,.
\ea
Next we assume $N$ is large enough that $1-e^{-iB_{\alpha
k}(\bmaths{\lambda})/N} \approx iB_{\alpha k}(\bmath{\lambda})/N$; then
\be
F_a = \sum_{\alpha =
0}^{M-1}a_\alpha e^{-iA_{\alpha}(\vec{\lambda})}\sum_{k=0}^{N-1} \tilde{x}_{\alpha
k}P[B_{\alpha k}(\bmath{\lambda})]
\ee
where
\ba
P[x] &=& \frac{1}{N}\frac{1-e^{-ix}}{1-e^{-ix/N}} \nonumber
\\ &\approx& \frac{\sin x}{x} -i\frac{1-\cos x}{x} \,.
\ea
Now arises the great advantage of the SFT method: the function $P[x]$ is sharply peaked about $x=0$,
so the sum over $k$
can be approximated by retaining only a few terms:
\be \label{eq:Fa-LALDemod}
F_a = \sum_{\alpha =
0}^{M-1}a_\alpha e^{-iA_{\alpha}(\vec{\lambda})}\sum_{k^\prime= -D}^{ D}
\tilde{x}_{\alpha k^\prime}P[B_{\alpha k^\prime}(\bmath{\lambda})]
\ee
where $k^\prime = k - k^\star$ and $k^\star$ is the value of $k$ such
that $B_{\alpha k^\star}(\bmath{\lambda}) = 0$, and $D$ is the number of
terms that we retain in the sum on either side of $k^\star$.
It turns out that $D=16$ suffices to calculate the
$\F$-statistic to within a few percent.

Eq.~(\ref{eq:Fa-LALDemod}) is our final approximation for $F_a$.
Analogous expressions hold for $F_b$, and the final $\F$-statistic
is calculated from $F_a$ and $F_b$ using Eq.~(\ref{eq:fstatdef}).
Thus, with the SFT method, for each point ${\bf \lambda}$ in parameter
space we need to calculate $A_\alpha(\vec{\lambda})$,
$B_{\alpha k}(\bmath{\lambda})$, and the amplitude modulation functions,
$a_\alpha$ and $b_\alpha$, and then to perform the sums in
Eq.~(\ref{eq:Fa-LALDemod}). It is then easy to see that to calculate the
$\F$-statistic for $n$ frequency bins, for a fixed value of
$\vec{\lambda}$, the number of floating point operations required is
roughly some $\textrm{constant}$ times $nMD$.   To estimate the constant,
let $C_1$ be the cost of calculating $P[B_{\alpha k}]$ for each
$\alpha$ and $k$ value.  Since multiplying two complex
numbers requires $6$ operations, and adding two complex numbers requires $2$
operations, we see that calculating the sum over $k^\prime$ in equation
(\ref{eq:Fa-LALDemod}) requires
\be (C_1 + 6)(2D + 1) + 4D
\ee
operations. Similarly, if $C_2$ is the cost of calculating $a_\alpha e^{-iA_{\alpha}}$ for
every $\alpha$, then the cost of calculating $F_a$ is
\be M[(C_1 + 6)(2D + 1) + 4D] + MC_2 + 6M + 2(M-1) \,.\ee
Thus, to find
$F_a$ and $F_b$ for every frequency bin requires $\approx 2\times
(2C_1 + 16)DM$ operations.  Since the cost of combining $F_a$ and $F_b$ to
get $\F$ is negligible compared to this, and assuming $C_1$ to be of
order unity, we see that the operation count for
calculating the $\F$-statistic
for $n$ frequency bins is approximately
\be
\sim 40 nMD\,.
\ee
For the first stage in the hierarchical search, the
$\F$-statistic is evaluated for $N^{(1)}N_{pc}^{(1)}f_\max\Tcoh$ bins.
So, taking $D=16$ and using $M = \Tcoh^{(1)}/\Tsft$, the cost is
\be \label{eq:sftcostapp}
\approx 640 N^{(1)}N_{pc}^{(1)}f_\max \frac{(\Tcoh^{(1)})^2}{\Tsft} \, .
\ee
\noindent
At higher stages, we evaluate the $\F$-statistic
\be
F^{(i-1)}
\textrm{max}\left\{1,\frac{N_{pc}^\ib}{N_{pf}^{(i-1)}} \right\}
\textrm{max}\left\{1,\frac{\Tcoh^\ib}{\Tcoh^{(i-1)}}\right\}\, N^{(i)}
\ee
times, so the operations count is
\ba
&&F^{(i-1)}
\textrm{max}\left\{1,\frac{N_{pc}^\ib}{N_{pf}^{(i-1)}} \right\}
\textrm{max}\left\{1,\frac{\Tcoh^\ib}{\Tcoh^{(i-1)}}\right\}  \nonumber \\
&& \ \ \times \left[\frac{640 N^\ib\Tcoh^\ib}{\Tsft} \right] \,.
\ea

\end{appendix}


\begin{thebibliography}{}

\bibitem{bccs}   P.R.~Brady, T.~Creighton, C.~Cutler, and B.F.~Schutz,
 \textit{Phys. Rev.} {\bf D57} 2101 (1998).

\bibitem{bc} P.R.~Brady and T.~Creighton,
\textit{Phys. Rev.} {\bf D61}, 082001 (2000).

\bibitem{jks}   P.~Jaranowski, A.~Kr\'olak, and B.F.~Schutz,
 \textit{Phys. Rev.} {\bf D58} 063001  (1998).

\bibitem{S1:pulsar} B.~Abbott et al. (The LIGO Scientfic
Collaboration),
\textit{Phys. Rev.} \textbf{D69}, 082004 (2004).

\bibitem{ip} R.~Prix and Y.~Itoh, \texttt{gr-qc/0504006}.

\bibitem{hough} B.~Krishnan, A.M.~Sintes, M.A.~Papa, B.F.~Schutz,
S.~Frasca, and C.~Palomba,
\textit{Phys. Rev.} \textbf{D70}, 082001 (2004).

\bibitem{lscsoft} The software can be found on the following websites:
  \texttt{http://www.lsc-group.phys.uwm.edu/daswg/projects/\\lal.html}
  and \texttt{http://www.lsc-group.phys.uwm.edu/\\daswg/projects/lalapps.html}

\bibitem{owen} B.~Owen, \textit{Phys. Rev.} \textbf{D53}, 6749 (1996).

\bibitem{jhc} J.H.~Conway and N.J.A.~Sloane, \textit{Sphere Packings,
Lattices and Groups}, Springer (1991).

\bibitem{abjk} P.~Astone, K.M.~Borkowski, P.~Jaranowski, and
A.~Kr\'olak,
\textit{Phys. Rev.} {\bf D65} 042003 (2003).

\bibitem{mrrtt} N.~Metropolis, A.~Rosenbluth, M.~Rosenbluth,
  A.~Teller, E.~Teller, \textit{J. Chem. Phys.} \textbf{21}, 1087
  (1953).

\bibitem{kgv} S.~Kirkpatrick, C.D.~Gelatt Jr., and M.P.~Vecchi,
  \textit{Science} \textbf{220}, No. 4598, 671 (1983).

\bibitem{nm} J.A.~Nelder and R.~Mead, \textit{Comput. J.} \textbf{7},
  308-313, 1965.

\bibitem{nr} W.H.~Press, \textit{Numerical Recipes in C}, Cambridge
  University Press (2002).

\bibitem{gsl} \texttt{http://www.gnu.org/software/gsl/}.

\bibitem{vecchio} S.V~Dhurandhar and A.~Vecchio,
 \textit{Phys. Rev.} {\bf D63} 122001 (2001).

\bibitem{eah}
  \texttt{http://www.physics2005.org/events/einsteinathome/\\\#einsteinathome}

\bibitem{emri}  J.\ R.\ Gair, L.\ Barack, T.\ Creighton,
C.\ Cutler, S.\ L.\ Larson, E.\ S.\ Phinney, and M.\ Vallisneri,
Proceedings of the Eighth GWDAW Meeting (Milwaukee, 2003); gr-qc/0405137.

\end{thebibliography}
\end{document}